\documentclass[prx,twocolumn,showpacs,superscriptaddress,preprintnumbers,amssymb]{revtex4-2}
\usepackage{graphicx}
\usepackage{latexsym}
\usepackage{amsmath}
\usepackage{amsfonts}
\usepackage{upgreek}
\usepackage{bm}
\usepackage{multirow}
\usepackage{enumitem}
\usepackage{color}
\usepackage[colorlinks, citecolor=blue]{hyperref}
\usepackage{physics}
\usepackage{tcolorbox}
\usepackage{manfnt}
\usepackage{relsize}

\newcommand{\uc}{{\hat{\rho}}}

\newcommand{\beq}{\begin{equation}}
\newcommand{\eeq}{\end{equation}}
\newcommand{\beqn}{\begin{eqnarray}}
\newcommand{\eeqn}{\end{eqnarray}}

\newcommand{\p}{\partial}

\newcommand{\cO}{ {\cal O} }

\newcommand{\cE}{ {\cal E} }

\newcommand{\cH}{ {\cal H} }
\newcommand{\cK}{ {\cal K} }
\newcommand{\cL}{ {\cal L} }
\newcommand{\cP}{ {\cal P} }

\newcommand{\bx}{\mathbf{x}}
\newcommand{\by}{\mathbf{y}}

\newcommand{\U}{\mathrm{U}}

\newcommand{\lind}{{\mathcal{L}}}

\newcommand{\bk}{{\bm{k}}}

\newcommand{\eqnref}[1]{Eq.\,\eqref{#1}}

\newcommand{\figref}[1]{Fig.\,\ref{#1}}

\newcommand{\rd}{\partial}

\newcommand{\vdagger}{{\vphantom{\dagger}}}

\newcommand{\sS}{{{\cal S}}}

\newcommand{\rAngle}{\rangle \hspace{-2pt} \rangle }
\newcommand{\lAngle}{\langle \hspace{-2pt} \langle }

\begin{document}

\title{Unifying Emergent Hydrodynamics and Lindbladian Low Energy Spectra across \\ Symmetries, Constraints, and Long-Range Interactions}

\author{Olumakinde Ogunnaike}
\affiliation{Department of Physics, Massachusetts Institute of Technology, Cambridge, MA 02139, USA}
\author{Johannes Feldmeier}
\affiliation{Department of Physics, Harvard University, Cambridge, MA 02138, USA}
\author{Jong Yeon Lee}\email{jongyeon@illinois.edu}
\affiliation{Kavli Institute for Theoretical Physics, University of California, Santa Barbara, CA 93106, USA}
\affiliation{Department of Physics, University of Illinois at Urbana-Champaign, Urbana, Illinois 61801, USA}

\date{\today}
\begin{abstract}
We identify emergent hydrodynamics governing charge transport in Brownian random time evolution with various symmetries, constraints, and ranges of interactions. This is accomplished via a mapping between the averaged dynamics and the low-energy spectrum of a Lindblad operator, which acts as an effective Hamiltonian in a doubled Hilbert space. By explicitly constructing dispersive excited states of this effective Hamiltonian using a single-mode approximation, we provide a comprehensive understanding of diffusive, subdiffusive, and superdiffusive relaxation in many-body systems with conserved multipole moments and variable interaction ranges. Our approach further allows us to identify exotic Krylov-space-resolved diffusive relaxation despite the presence of dipole conservation, which we verify numerically.
Therefore, we provide a general and versatile framework to qualitatively understand the dynamics of conserved operators under random unitary time evolution. 
\end{abstract}

\maketitle

\emph{Introduction.}
Recent years have seen a surge of interest in the nonequilibrium dynamics of quantum many-body systems, driven by rapid advancements in quantum simulation capabilities across diverse physical platforms.  
In particular, significant attention has been devoted to understanding the thermalization process of interacting many-body systems~\cite{Deutsch1991, Srednicki1994, Rigol2008, Polkovnikov2011, Alessio2016,kaufman2016_thermalization,Brydges2019_renyi}. A vital theoretical tool that provides key insights into the dynamics of thermalizing quantum systems is the study of random unitary time evolution. While retaining analytical tractability, such methods can successfully capture universal properties of non-integrable many-body dynamics such as transport, operator spreading, or entanglement growth~\cite{Nahum2017, Keyserlingk2018, Hamma2012, Hamma2012b, Pavel2018, Gharibyan2018, Moudgalya2019, Vedika2018, Rakovszky2018}.
In particular, the application of methods based on random unitary evolution has highlighted the importance of symmetries and constraints in many-body dynamics, unveiling a rich phenomenology of emergent hydrodynamics at late times. Recent results range from transport in long-range interacting systems~\cite{Schuckert2020LR,Joshi2022LR,richter2023_long,zu2021emergent} to anomalously slow subdiffusion~\cite{Iaconis19, guardado2020subdiffusion, gromov2020fracton, feldmeier2020anomalous, morningstar2020kinetically,zhang_2020,Moudgalya2021, Singh2021_subdiff,feldmeier2021_fractondimer,sala2022_modulated,hart2022_hidden,feldmeier2022_tracer} or even localization due to Hilbert space fragmentation in models with kinetic constraints~\cite{Sala2020, Khemani2020_shattering, Rakovszky2020_fragmentation, Moudgalya2019_arxiv, Yang2020, tomasi2019, Roy2020, Herviou2021, Shibata2020, Langlett2021,scherg2021_kinetic,feldmeier2021_crit}.

\begin{figure}\label{fig:Lindbladian}
    \includegraphics[width=0.49\textwidth]{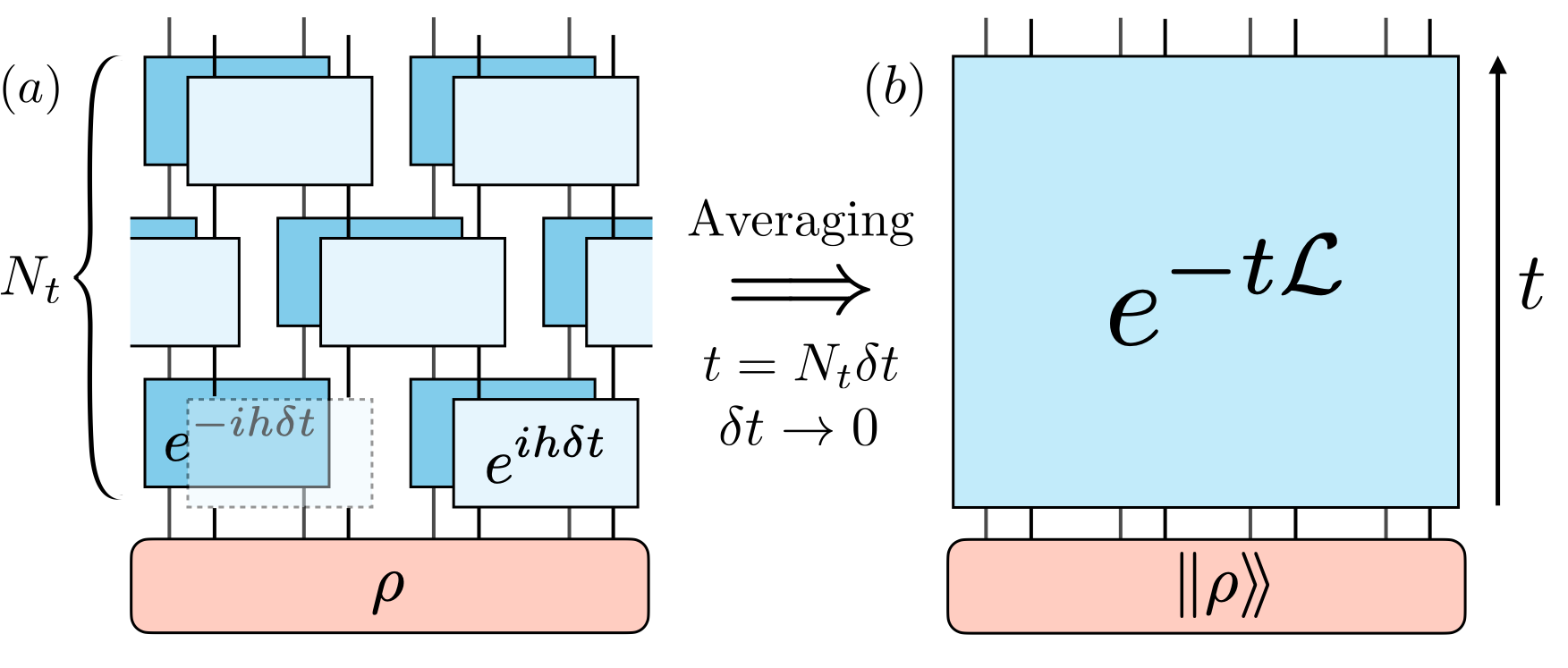}
    \caption{ {\bf Brownian circuit and effective Hamiltonian.} Mapping (a) random operator dynamics to (b) imaginary-time evolution by an effective Hamiltonian $\lind$ in a doubled Hilbert space. On the left, an operator $\rho$ is evolved by a local Hamiltonian $H_t$\,$\equiv$\,$\sum_i h_i dB_{i,t}$ with Brownian random variable $dB$.  Overlapping blocks for forward/backward evolution (dark/light) share the same Brownian variable, but all other Brownian variables are independently drawn from Gaussian distributions. On the right, we average over random variables while taking timesteps to zero; this produces imaginary-time Schrodinger evolution by a Lindbladian operator.}
\end{figure}

In this work, we introduce a simple, yet powerful method to understand the qualitative behavior of late-time hydrodynamics based on Brownian Hamiltonian evolution, which can be modeled by Markovian dynamics and thus captured by a Lindblad equation ~\cite{CALDEIRA1983587,  Buchleitner_Hornberger_Kümmerer_2011, Lashkari2013,  ZhouXiao2019, XuSwingle2019, XiaoZhou2019}.
Our approach successfully reproduces results reported in previous literature and allows us to uncover novel, unconventional hydrodynamic relaxation in constrained many-body systems.
The key technical step relates dynamical properties such as the auto-correlation of conserved operators to the low-energy spectrum of an emergent effective Hamiltonian in a \emph{doubled Hilbert space}~\cite{CHOI1975,JAMIOLKOWSKI1972}. 
The low-energy excitation spectrum of the latter thus dictates the long-time dynamics of such correlations. Accordingly, this mapping allows us to utilize well-established techniques in condensed matter physics, such as the single-mode approximation, to analyze our problem. 
Here, we apply this method to various scenarios: We show that systems conserving $\U(1)$ global charge as well as higher multipole moments exhibit diverse hydrodynamic relaxation depending on their symmetries and ranges of interactions. Then, we extend our approach to understand Krylov-subspace-resolved hydrodynamics, where we uncover general conditions under which relaxation is diffusive despite the presence of dipole conservation. We verify this diffusive relaxation numerically in lattice models in both one and two spatial dimensions. 

\emph{Brownian Circuits}. We consider time evolution by a time-dependent Hamiltonian $H_t$\,$\equiv$\,$\sum_i h_i\, dB_{i,t}$, defined via interaction terms $h_i$ with $h_{\boldsymbol{x},\lambda}\,{=}\,h_{\boldsymbol{x} + \boldsymbol{1},\lambda}$ and Brownian random variables $ d B_{t,i}$ at each time slice $[t,t+\delta)$. Here, the label $i$\,$=$\,$(\boldsymbol{x},\lambda)$ encodes both the spatial support and operator type of $h_i$. The random variables have vanishing mean $\mathbb{E}[dB]$\,$=$\,$0$ and finite variance $\mathbb{E}[dB^2]$\,$=$\,$1/\delta$. 

Under this time evolution, a density matrix  $\rho(t)$ evolves as $\rho(t+\delta)$\,$\equiv$\,$e^{-i H_t \delta} \rho(t) e^{i H_t \delta}$.
Averaging the infinitesimal time evolution over the random variables, the leading order operator evolution becomes~\cite{sm}(Sec.\,B):
\begin{align} \label{eq:dynamics}
   \mathbb{E}[ \rd_t \rho ] &=- \frac{1}{2} \sum_i ( h_i^2 \rho - 2 h_i \rho h_i + \rho h_i^2)  =  \cL[\rho],
\end{align}
where ${\cL}$ is a superoperator called the Lindblandian. 

We now construct an alternative description of the operator dynamics \eqnref{eq:dynamics} by employing the Choi isomorphism, a mapping from an operator acting on the Hilbert space ${\cal H}$ to a state defined on the doubled Hilbert space ${\cal H}_u \otimes {\cal H}_l$, where subscripts $u,l$ are introduced to distinguish two copies of $\cH$. For a given operator $O$, the mapping reads $O$\,\,$\mapsto$\,\,$\Vert O \rAngle$\,$\equiv$\,$\sum_i |i \rangle \otimes \big( O|i \rangle \big)$,
where the summation is over all basis states of the original Hilbert space~\cite{sm}(Sec.\,A). Under this mapping, the Lindbladian superoperator $\lind$ maps to a linear operator $\hat{ {\cal H} }_{\cal L}$ acting on the doubled Hilbert space:
\begin{align} \label{eq:Lindbladian}
    \hat{ {\cal H} }_{\cal L} &= \frac{1}{2} \sum_i \qty|h_i^T \otimes \mathbb{I} -  \mathbb{I} \otimes h_i |^2=: \frac{1}{2}\sum_{{\bm{x}},\lambda} {\cal O}^\dagger_{{\bm{x}},\lambda}  {\cal O}^\vdagger_{{\bm{x}},\lambda}.
\end{align}
where $|\dots|^2$ should be understood as $(\dots)^\dagger(\dots)$, and ${\cal O}^\vdagger_{{\bm{x}},\lambda} = (h_{{\bm{x}},\lambda}^T \otimes \mathbb{I} -  \mathbb{I} \otimes h_{{\bm{x}},\lambda})$.  The average dynamics in \eqnref{eq:dynamics} can then be recast into an imaginary time evolution generated by the effective Hamiltonian $\hat{ {\cal H} }_{\cal L}$: 
\begin{align} \label{eq:brownian1}
    \rd_t \Vert O \rAngle=- \hat{ {\cal H} }_{\cal L} \Vert O
 \rAngle \quad \Rightarrow \quad \Vert O(t) \rAngle=e^{- t \hat{ {\cal H} }_{\cal L} } \Vert O_0 \rAngle.
\end{align}
 
We are interested in the dynamics of a local operator $O$ under Brownian evolution, which we characterize by the averaged auto-correlation function $\mathbb{E}  \langle O_{{\bm{y}}}(0) O_{{\bm{x}}}(t) \rangle_\rho$~\footnote{It measures the spreading of an operator $O$ located at $\bx$ at time $t$ by measuring its overlap with an operator at $\by$.} with respect to the maximally mixed state $\rho$\,$=$\,$\frac{1}{D} \mathbb{I}$, where $D$ is the dimension of the many-body Hilbert space.

Note that \eqnref{eq:Lindbladian} inherits translation invariance from the interaction terms, $h_{\boldsymbol{x},\lambda} = h_{\boldsymbol{x} + \boldsymbol{1},\lambda}$. Therefore, we can label the eigenstates of $\hat{ {\cal H} }_{\cal L}$ by their momentum; let $\Vert {\bm{k}}, \nu \rAngle$ be the eigenstates of $\hat{ {\cal H} }_{\cal L}$ with energy $E_{\boldsymbol{k},\nu}$, carrying momentum $\bk$ and an additional label $\nu$. Inserting a completeness relation, we obtain
\begin{align}\label{eq:decompose}
     & \mathbb{E}  \langle O_{{\bm{y}}}(0) O_{{\bm{x}}}(t) \rangle_\rho =\frac{1}{D}  \lAngle O_{\bm{y}}(0) \Vert e^{-t \hat{ {\cal H} }_{\cal L}} \Vert O_{\bm{x}}(0) \rAngle \nonumber \\
    &\qquad \qquad =\frac{1}{D} \sum_{{\bm{k}},\nu} e^{-t E_{{\bm{k}},\nu}} e^{i{\bm{k}} \cdot ({\bm{y}}-{\bm{x}})} | \lAngle {\bm{k}},\nu \Vert O_{{\bm{x}}} \rAngle|^2.    
\end{align}
Consider a $d$-dimensional system. Assuming a gapless dispersion $\min_{\nu}\{E_{{\bm{k}},\nu}\}$\,$\sim$\,$k^n$ at low momentum $\bk$\,$\rightarrow$\,$0$, as well as a finite overlap $|\lAngle \bk,\nu \Vert O_x \rAngle|^2$ of the operator of interest $\Vert O_{\bm{x}} \rAngle$ with these gapless modes~\cite{misc2}, the autocorrelation  at $x\,{=}\,y$ decays algebraically as 
\begin{align}
    \mathbb{E}  \langle O_{{\bm{x}}}(t) O_{{\bm{x}}}(0)\rangle_\rho  \underset{t \rightarrow \infty}{\sim}  \int_k e^{-t k^n} \dd^d \bk \sim t^{-d/n},
\end{align}
implying that the dynamical exponent $z$\,$=$\,$n$. 
Therefore, the study of late-time operator dynamics in the Brownian evolution reduces to the identification of gapless dispersing states in the effective Hamiltonian $\hat{\mathcal{H}_{\lind}}$.

\emph{Charge Conservation.} We now assume that each $h_i$ in the original Hamiltonian exhibits a $\U(1)$ charge conservation symmetry. 
In the doubled Hilbert space, the symmetry is doubled as well, and the effective Hamiltonian $\hat{ {\cal H} }_{\cal L}$ in \eqnref{eq:Lindbladian} must be symmetric under $G$\,$=$\,$\U(1)_u \times \U(1)_l$. We denote by $G_\textrm{diag}$ and $G_\textrm{off}$ the diagonal and off-diagonal subgroups of $G$, generated by $g_{\textrm{diag/off}}$\,$=$\,$\hat{Q}_u \otimes \mathbb{I} \mp \mathbb{I}\otimes \hat{Q}_l$, where $\hat{Q}$ is the total charge operator~\cite{sm}(Sec.\,A).

First, we examine the ground states of $\hat{ {\cal H} }_{\cal L}$, which is  positive semidefinite. The Choi state of the identity operator $\Vert \mathbb{I} \rAngle$ satisfies $\hat{ {\cal H} }_{\cal L} \Vert \mathbb{I} \rAngle$\,$=$\,$0$ and is thus a ground state of $\hat{ {\cal H} }_{\cal L}$. Due to $\U(1)$ symmetry, $\mathbb{I}$ decomposes into the summation over projectors onto different charge sectors: $\mathbb{I}$\,$=$\,$\sum_m \cP_m$, where $\cP_m$ is the projector onto a $\U(1)$ sector of charge $m$. For a system with $N = L^d$ sites and local Hilbert space dimension $M$, $m \in \{ 0,1,..., M L^d\}$. We denote $\Vert m \rAngle$ as the Choi state of $\cP_m$. As such, $\Vert m \rAngle$ is also a ground state of $\hat{ {\cal H} }_{\cal L}$ with vanishing $G_\textrm{diag}$ charge and a $G_\textrm{off}$-charge of $2m$. Note that $\lAngle m \Vert m \rAngle$\,$=$\,$\dim[\cH_m]$, the dimensionality of the charge-$m$ sector. Moving forward, we renormalize $\Vert {m} \rAngle$ to $\lAngle {m} \Vert {m} \rAngle$\,$=$\,$1$. 

The degenerate groundstate manifold with different $G_\textrm{off}$-charges implies spontaneous symmetry breaking of $G_\textrm{off}$. This can be shown explicitly by constructing a groundstate state $\Vert \theta \rAngle \equiv \sum_{m} f(m) e^{im \theta} \Vert m \rAngle$ such that under the rotation by $G_\textrm{off}$ generator, $e^{i \alpha g_{\textrm{off}}} \Vert \theta \rAngle  = \Vert \theta + \alpha \rAngle \neq \Vert \theta \rAngle$. The low-energy excitations of $\hat{\mathcal{H}_{\lind}}$ must be given by the Nambu-Goldstone modes for the broken continuous symmetry. A standard approach for constructing  Goldstone modes is to apply $G_\textrm{off}$ density modulations with momentum $k$ on the ground state $\Vert m \rAngle$.  
The variational ansatz for such a state is defined as
\begin{align} \label{eq:density_twist}
    \Vert m_{{\bm{k}}} \rAngle &\equiv \frac{1}{ \sqrt{\mathcal{N}_{\boldsymbol{k}}} } \uc_{{\bm{k}}} \Vert {m} \rAngle, \quad \uc_{{\bm{k}}}  \equiv   \sum_{{\bm{x}}} \frac{e^{i{\bm{k}} \cdot  {\bm{x}}}}{L^{d/2}} ( \uc_{{\bm{x}},u} + \uc_{{\bm{x}},l}),
\end{align}
where $\uc_{{\bm{x}},u/l}$ measures $\U(1)$ charge in the layer $u$ or $l$ at position ${\bm{x}}$, and ${\cal N}_{{\bm{k}}}$\,$\equiv$\,$\lAngle {m} \Vert \uc_{{\bm{k}}}^\dagger \uc^\vdagger_{{\bm{k}}} \Vert {m} \rAngle$ is a static structural factor with $ \uc_{{\bm{k}}}^\dagger$\,$=$\,$\uc_{-{\bm{k}}}^\vdagger$. 
It straightforward to show that $\Vert m_{{\bm{k}}} \rAngle$ carries a well-defined momentum ${\bm{k}}$ and thus $\lAngle m_{{\bm{k}}} \Vert m_{{\bm{k}}'} \rAngle$\,$=$\,$\delta_{{\bm{k}},{\bm{k}}'}$~\cite{sm}(Sec.\,E). 
We remark that since $( \uc_{{\bm{x}},u} + \uc_{{\bm{x}},l})$ measures a local $G_\textrm{off}$-charge, the constructed mode corresponds to the density fluctuations of the $G_\textrm{off}$-charge.

What is the energy of this variational state? With orthogonality between $\Vert m_\bk \rAngle$ for different momenta, the variational expected energy provides an upper bound for the low-energy dispersion of \eqnref{eq:Lindbladian}:
\begin{align} \label{eq:fdr}
    &\lAngle m_{{\bm{k}}} \Vert \hat{ {\cal H} }_{\cal L} \Vert m_{{\bm{k}}} \rAngle= \frac{1}{ {\cal N}_{{\bm{k}}} } \sum_{{\bm{x}},\lambda} \lAngle {m} \Vert [{\cal O}_{{\bm{x}}, \lambda}, \uc_{{\bm{k}}}]^\dagger [{\cal O}_{{\bm{x}}, \lambda}, \uc_{{\bm{k}}}] \Vert {m} \rAngle,
\end{align}
where we used ${\cal O}_{{\bm{x}},\lambda} \Vert m \rAngle$\,$=$\,$0$. By using $\U(1)$ symmetry, the commutator in \eqnref{eq:fdr} can be recast as
\begin{align} \label{eq:commutator}
    [{\cal O}_{{\bm{x}}, \lambda}, \uc_{{\bm{k}}}] 
    = e^{i\boldsymbol{k}\cdot \boldsymbol{x}}  \sum_{\boldsymbol{y} \in \sS_{{\bm{x}}}} \sum_{n=1}^{\infty} [{\cal O}_{{\bm{x}}, \lambda}, \frac{[i\boldsymbol{k}\cdot(\boldsymbol{y}-\boldsymbol{x})]^n}{n!} \uc_{\boldsymbol{y}} ],
\end{align}
where we used $[\cO_{\boldsymbol{x},\lambda}, \sum_{\boldsymbol{y}} \uc_{\boldsymbol{y}}]$\,$=$\,$0$, and $\sS_{{\bm{x}}}$ is the \textit{local} support of the operator ${\cal O}_{{\bm{x}},\lambda}$ (thus warranting the expansion of $e^{i\boldsymbol{k}\cdot(\boldsymbol{y}-\boldsymbol{x})}$ for small $\boldsymbol{k}$). Generally, assuming a finite expectation value of the local dipole fluctuations $\lAngle {m}\Vert \bigl|[{\cal O}_{{\bm{x}}, \lambda}, \sum_{\boldsymbol{y}} y_{i}\, \uc_{\boldsymbol{y}} ]\bigr|^2 \Vert {m} \rAngle$, the expansion \eqnref{eq:commutator} does not vanish at $n$\,$=$\,$1$, giving rise to a leading order contribution proportional to $k$:
\begin{align}
     [{\cal O}_{{\bm{x}}, \lambda}, \uc_{{\bm{k}}}] \propto k \quad \Rightarrow \quad  \lAngle m_{{\bm{k}}} \Vert \hat{ {\cal H} }_{\cal L} \Vert m_{{\bm{k}}} \rAngle \propto k^2.
\end{align}
Here, we focus on isotropic systems for simplicity; however, dynamical exponents can be obtained similarly for non-isotropic systems. Furthermore, ${\cal N}_{{\bm{k}}}$ is a constant, independent of $k$~\cite{sm}(Sec.\,E). Therefore, $\Vert m_{{\bm{k}}} \rAngle$ generically exhibits a quadratic ($E_{{\bm{k}}}$\,$\propto$\,$k^2$) dispersion, regardless of the details of the effective Hamiltonian.
Note the similarity of our approach to the single-mode approximation in superfluid or quantum Hall states~\cite{Feynman1953, Feynman1954, Girvin1986}, where the Feynman-Bijl ansatz~\cite{sm}(Sec.\,D) provides variational states that capture the dispersion of density fluctuation excitations. 

\emph{Long-Range Interactions}. We extend our preceeding analysis to charge-conserving systems with long-range interactions. Specifically, we consider the effects of long-range terms in our Hamiltonian of the form $h_{{\bm{x}}, {\bm{x}}'}= |{\bm{x}}- {\bm{x}}'|^{-\alpha} (\hat{S}^+_{{\bm{x}}}\hat{S}^-_{{\bm{x}}'} + \textrm{h.c.})$,
where $\hat{S}^\pm_{{\bm{x}}}$ are raising and lowering operators for the charge $\uc_{\boldsymbol{x}}$ at site ${\bm{x}}$ and $\hat{Q}=\sum_x \uc_{\boldsymbol{x}}$ is conserved. The effective Hamiltonian reads $\hat{ {\cal H} }_{\cal L}=\sum_{{\bm{x}},{\bm{x}}'} {\cal O}_{{\bm{x}}, {\bm{x}}'}^\dagger {\cal O}_{{\bm{x}}, {\bm{x}}'}^\vdagger$ and the commutator entering \eqnref{eq:fdr} becomes
\begin{align}\label{eq:longrange_commutator}
    [{\cal O}_{{\bm{x}}, {\bm{x}}'}, \uc_{{\bm{k}}}] =e^{i {\bm{k}} \cdot {\bm{x}}}\frac{(1 - e^{i{\bm{k}} \cdot({\bm{x}}'-{\bm{x}})})}{|{\bm{x}}-{\bm{x}}'|^{\alpha} }\left[\tilde{{\cal O}}_{{\bm{x}}, {\bm{x}}'},  \uc_{{\bm{x}}} \right],
\end{align}
where $\tilde{{\cal O}}_{{\bm{x}}, {\bm{x}}'}$\,$:=$\,${\cal O}_{{\bm{x}}, {\bm{x}}'}|{\bm{x}}$\,$-$\,${\bm{x}}'|^{\alpha}$ is now distance-independent.
Assuming $\alpha$\,$>$\,$d/2$ and a finite expectation value for the square of the commutator on the RHS of \eqnref{eq:longrange_commutator}, the variational energy of $\Vert m_{{\bm{k}}} \rAngle$ is~\cite{sm}(Sec.\,F)
\begin{align} \label{eq:longrange_charge}
\lAngle m_{{\bm{k}}} \Vert \hat{ {\cal H} }_{\cal L} \Vert m_{{\bm{k}}} \rAngle \underset{k \rightarrow 0}{\propto}  C_1(\alpha) |k|^{2\alpha-d} + C_2(\alpha)k^2 .
\end{align}
Thus, for $\alpha$\,$<$\,$1$\,$+$\,$d/2$, the system relaxes superdiffusively with $z$\,$=$\,$2 \alpha - d$, successfully reproducing previous works on long-range interacting systems~\cite{Schuckert2020LR, Joshi2022LR, deng2022superdiffusion}.
Alternatively, for $\alpha$\,$\leq$\,$d/2$ the prefactors $C_1(\alpha)$ and $C_2(\alpha)$ exhibit divergences and the associated modes become gapped~\cite{sm}(Sec.\,F); accordingly, the operator decays exponentially fast~\cite{Schuckert2020LR}, entering an effectively nonlocal ``all-to-all'' interacting regime. 
 

\emph{Dipole Conservation}. The method outlined above also applies to systems with conserved quantities beyond $U(1)$ charges. 
Let us focus on one-dimensional models with charge multipole symmetries, as relevant to fracton systems~\cite{nandkishore2019_fractons,pretko2020fracton,chamon2005_glass,Haah11,yoshida2013_fractal,vijay2015_topo,pretko2018_elasticity,PretkoSub,pretko2018_gaugprinciple}, generated by $Q^{(n)} \equiv \sum_{x} x^n \uc_{x}=\sum_{x} x^n (\uc_{x,u} + \uc_{x,l})$.
Concretely, we consider Brownian time evolution conserving the first two multipole moments $n$\,$=$\,$0$ and $n$\,$=$\,$1$, i.e. $[h_i,Q^{(0)}]$\,$=$\,$[h_i,Q^{(1)}]$\,$=$\,$0$. This combination of charge and dipole symmetries generally leads to Hilbert space fragmentation~\cite{Sala2020,Khemani2020_shattering,Moudgalya2022}: 
For a given symmetry sector $Q^{(0)},Q^{(1)}$ labeled by the different charge and dipole values, there are numerous distinct Krylov sectors, $\cK$, connected by the Hamiltonian evolution. 
Our goal is to understand the associated \textit{Krylov-space-resolved hydrodynamics} in such systems.
For this purpose, we introduce the operator ${\cal P}_{\mathcal{K}}$ projecting onto an individual Krylov sector, $\cK$, and its Choi state $\Vert \cK \rAngle$, which we define to be normalized. In the doubled Hilbert space formalism, we thus define new excited states, $\Vert \mathcal{K}_k \rAngle$\,$=$\,$\uc_k \Vert \mathcal{K} \rAngle/ (\mathcal{N}_k^{\cK})^{1/2}$, where $\hat{ {\cal H} }_{\cal L}\Vert \mathcal{K} \rAngle$\,$=$\,$0$ and ${\cal N}^\mathcal{K}_k$\,$\equiv$\,$\lAngle \mathcal{\cK} \Vert \uc_k^\dagger \uc_k^\vdagger \Vert \mathcal{K} \rAngle$ is the Krylov-resolved structure factor.

In the presence of both charge and dipole conservation symmetries, the commutator in \eqnref{eq:commutator} now vanishes at $n$\,$=$\,$1$, and takes a finite value only at order $n$\,$\geq$\,$2$. Accordingly, the excited modes $\Vert \mathcal{K}_k \rAngle$ carry an energy $E_k=\lAngle \mathcal{K}_k \Vert \hat{ {\cal H} }_{\cal L} \Vert \mathcal{K}_k \rAngle \propto \frac{1}{\mathcal{N}^\mathcal{K}_k} \, k^4$.
For generic dipole-conserving systems featuring \textit{weak} fragmentation, the largest Krylov sector $\mathcal{K}_0$ makes up a finite portion of the full Hilbert space (up to a prefactor algebraic in system size). As a consequence, its static structure factor $\mathcal{N}^{\mathcal{K}_0}_k$\,$\rightarrow$\,$\mathcal{O}(1)$ remains finite as $k$\,$\rightarrow$\,$0$. We thus obtain subdiffusive relaxation with dynamical exponent $z$\,$=$\,$4$. The generalization of this result to systems conserving $\{Q^{(0)},...,Q^{(m)}\}$ multipoles is straightforward: The commutator in \eqnref{eq:commutator} now vanishes up to order $n$\,$=$\,$m$, giving rise to a dispersion proportional to $k^{2(m+1)}$ and dynamical exponent $z$\,$=$\,$2(m+1)$, in accordance with previous 
results~\cite{gromov2020fracton,feldmeier2020anomalous,
morningstar2020kinetically,guardado2020subdiffusion,iaconis2021_multipole,Singh_2021}.

\begin{figure}
    \includegraphics[width=0.45\textwidth]{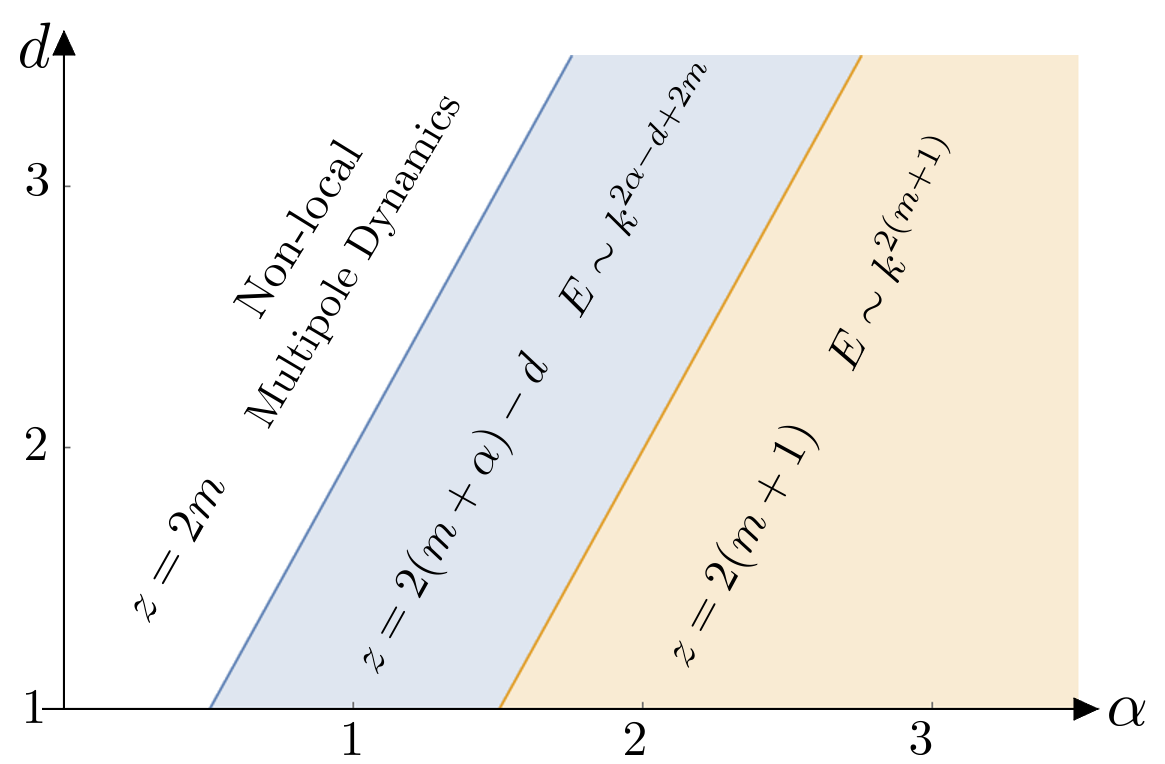}
    \caption{ \label{fig:pd} {\bf Relaxation dynamics in multipole-conserving systems with long-range interactions.} Systems with $\frac{1}{r^\alpha}$ power-law decaying hopping of local multipoles of order $m$ exhibit three distinct dynamical regimes.  When $\alpha$\,$>$\,$\frac{d}{2} + 1 $ (orange), the dynamics is (sub)diffusive with dynamical exponent $z$\,$=$\,$2(m+1)$.  For $\frac{d}{2}$\,$+$\,$1$\,$>$\,$\alpha$\,$>$\,$\frac{d}{2}$ (blue), the dynamics is faster, with dynamical exponent $z$\,$=$\,$2(m+\alpha)$\,$-$\,$d$. When $\alpha$\,$\leq$\,$\frac{d}{2}$, the system is effectively non-local, thus, relaxation occurs from individual $m$-th multipole creation/annihilation operators, which are hoppings of $(m-1)$-th multipole charges. This results in (sub)diffusive transport with $z$\,$=$\,$2(m-1)$\,$+$\,$2$\,$=$\,$2m$.
    }
    \label{fig:diffuse_phase}
\end{figure}

Similar to the charge-conserving case, these results can be extended to long-range interacting systems in arbitrary dimensions. For example, consider power-law decaying dipole hopping terms $h_{{\bm{x}},{\bm{x}}'}$\,$\sim$\,$\frac{1}{|{\bm{x}}-{\bm{x}}'|^\alpha} (D^\dagger_{{\bm{x}}}D^\vdagger_{{\bm{x}}'}$\,$+$\,$\textrm{h.c.})$, where $D_{{\bm{x}}}$ is a \textit{local} operator lowering the dipole moment.  When $\alpha$\,$>$\,$\frac{d}{2}$, we determine the dispersion to be $E_{{\bm{k}}}$\,$\sim$\,$C_1(\alpha) k^{2\alpha + 2 - d}$\,$+$\,$C_2(\alpha) k^4$~\cite{sm}(Sec.\,F). Therefore, if $\alpha$\,$<$\,$1$\,$+$\,$d/2$, charge spreads faster than the subdiffusive transport $z$\,$=$\,$4$ of short-range systems. 
For $\alpha$\,$<$\,$\frac{d}{2}$, 
dipole hopping becomes highly non-local, and charge transport effectively arises from individual local dipole creation/annihilation terms, analogous to systems with conventional charge conservation. 
In our framework, after renormalizing the single-mode dispersion to be bounded~\cite{sm}(Sec.\,F), we obtain $E_{{\bm{k}}}$\,$\sim$\,$k^2$. We provide a summary of the dynamical exponents emerging in multipole-conserving systems with such long-range hopping of local moments in \figref{fig:pd}.

\emph{Constrained dynamics.}
Returning to short-range models with dipole-conservation, we may ask whether relaxation differing  from the subdiffusive behavior $z=4$ can emerge in specific Krylov sectors. The presence of the structure factor in the dispersion of \eqnref{eq:fdr} suggests this may be the case in Krylov sectors where charge fluctuation follow a \textit{sub-volume} law with vanishing $\lim_{k\rightarrow 0}\mathcal{N}^\mathcal{K}_k=0$. We demonstrate this effect in concrete examples below.

Let us first consider a one-dimensional chain with charge and dipole conservation and introduce bond variables $\hat{e}_x$ defined via $\uc_x=\hat{e}_{x}-\hat{e}_{x-1}$, i.e. $\hat{e}_x=\sum_{i=0}^{x} \uc_i$.
For convenience, we define the charge density $\uc_x$ relative to its average value within $\mathcal{K}$, i.e. $\sum_x \left<\uc_x \right>_\mathcal{K}= 0$. We note that the $\hat{e}_i$ can be understood as a local dipole density, with $\sum_x \hat{e}_x=Q^{(1)}$~\cite{Moudgalya2021,skinner2022fracton,zechmann2022fractonic}.
Let us now assume that a sector $\mathcal{K}$ exhibits \textit{bounded} fluctuations of these bond variables. Formally, $\lim_{L \rightarrow \infty} \left< \hat{e}_k \hat{e}_{-k} \right>_{\mathcal{K}} \xrightarrow{k\rightarrow 0} \sigma_1^2 < \infty$,
where $\hat{e}_k=\frac{1}{\sqrt{L}}\sum_x e^{ikx} \hat{e}_x$ and $\sigma_1$ corresponds to the average fluctuation of the local dipole density. Since $\hat{e}_x=\sum_{i=0}^x \uc_i$, the finiteness of $\hat{e}_x$ implies area-law fluctuations of the total charge within any given region.
Using that $\uc_k$\,$=$\,$(1-e^{-ik})\hat{e}_k$ for $k$\,$\neq$\,$0$, the structure factor for small $k$ becomes
\begin{align} \label{eq:dipole5}
\mathcal{N}^{\mathcal{K}}_k=\left< \uc_k \uc_{-k}\right>_{\mathcal{K}}=k^2 \left< \hat{e}_k\hat{e}_{-k} \right> \rightarrow \sigma_1^2 \, k^2.
\end{align}
Therefore, for Krylov sectors satisfying \eqnref{eq:dipole5}, the energy of the excited mode $\Vert \cK_k \rAngle$ scales as $E_k$\,$\propto$\,$k^2$ and we expect \textit{diffusive} relaxation, despite the presence of dipole-conservation. 
To interpret this result, note that the $\hat{e}_x$ constitute a conserved local density with an effectively finite local state space due to their bounded fluctuations. If $\hat{e}_x$ is bounded, these local dipoles move without additional kinetic constraints and are thus expected to relax diffusively, see also Ref.~\cite{skinner2022fracton}. 
Generalization to systems conserving $\{Q^{(0)},...,Q^{(m)}\}$ is again straightforward: Krylov sectors with bounded multipole densities up to order $p$\,$\leq$\,$m$ have $\mathcal{N}_k$\,$\rightarrow$\,$\sigma_p^2 \, k^{2p}$, leading to a dispersion $\propto k^{2(m-p+1)}$ in short-range systems.

As a concrete example of \eqnref{eq:dipole5}, we consider random Brownian evolution in a $S=1$ spin chain with local dipole-conserving terms $h_i$\,$=$\,$\hat{S}_i^+ (\hat{S}_{i+1}^-)^2 \hat{S}_{i+2}^+$\,$+$\,$\textrm{h.c.}$. Although these terms induce a strong fragmentation of the Hilbert space, there exist exponentially large, delocalized Krylov sectors~\cite{Sala2020, Rakovszky2020_fragmentation}.
We label the local charge density by $\uc_x = S^z_x \in \{0, \pm \}$ and consider the Krylov sector containing the initial state $\ket{\psi_0}=\ket{...00+00...}$. 
In terms of the variables $\hat{e}_x$ introduced above, $\ket{\psi_0}=\ket{...00111...}$ corresponds to a domain wall, and the $\hat{e}_x \in \{0,1\}$ can be shown to take values in a bounded range~\cite{Sala2020}, thus satisfying our condition \eqnref{eq:dipole5}.
Diffusive relaxation of this state has indeed been found in Ref.~\cite{skinner2022fracton}, and $\mathbb{E} \langle S^z_{x=L/2}(t)\rangle$\,$\sim$\,$t^{-1/2}$ can be verified numerically using random classical time evolution, as illustrated in ~\cite{sm}(Sec.\,G).


\begin{figure}
\centering
\includegraphics[width=\columnwidth]{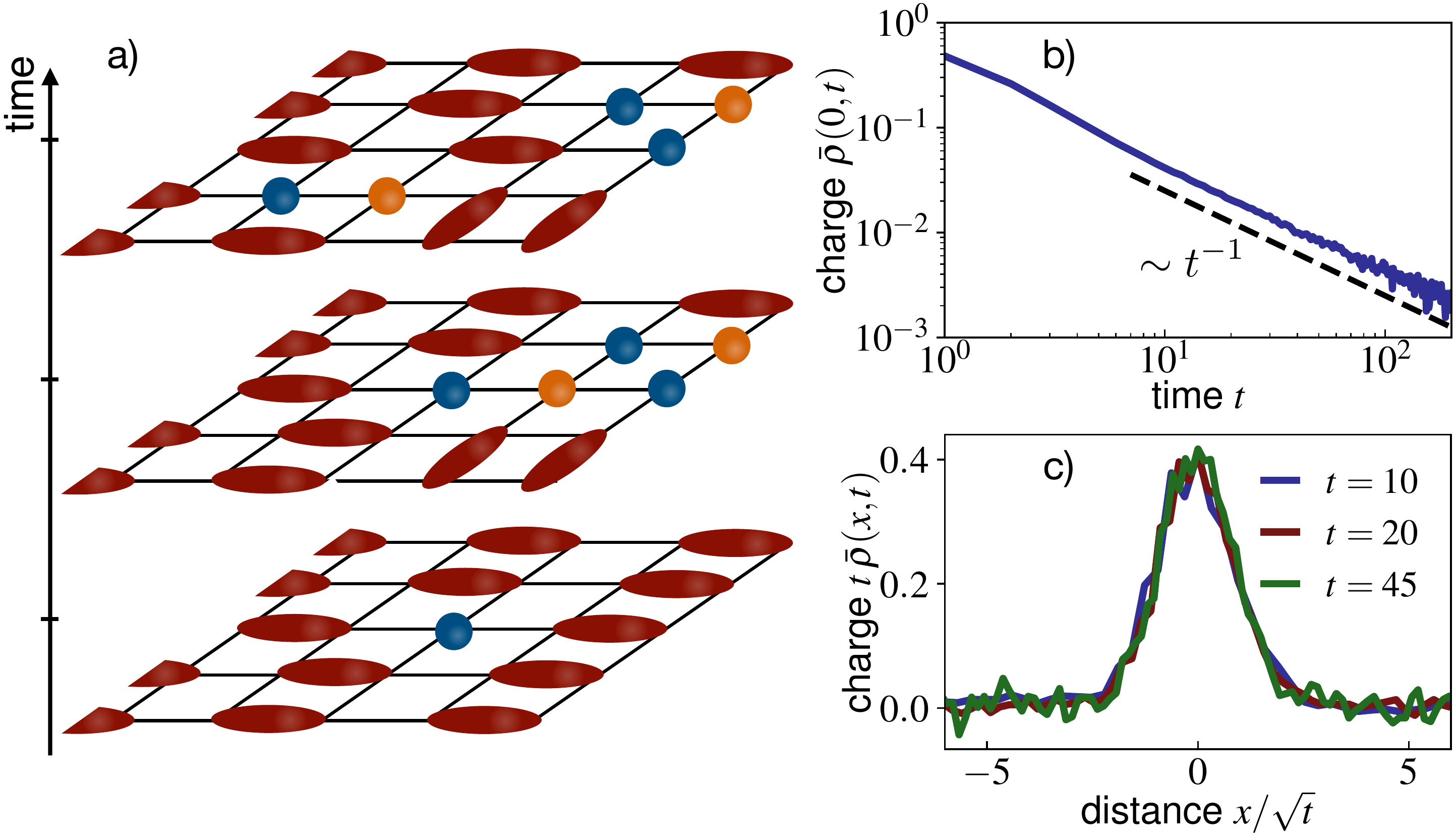}
\caption{\textbf{Relaxation dynamics in a dipole-conserving dimer model.} \textbf{a)} We numerically consider a classical, discrete random time evolution in a dimer model with hard-core constraint, i.e. maximally one dimer attached to each site in the square lattice. This model can be mapped onto a $U(1)$ link model following Refs.~\cite{horn1981finite,orland1990lattice,chandrasekharan1997quantum}. Under this mapping, vacancies, i.e. sites without attached dimer, carry positive (blue spheres) or negative charge (orange spheres), depending on their sublattice. 
We explicitly incorporate preservation of the hard-core constraint, the total charge, and the dipole moment associated with these charges in the time evolution. \textbf{b)} Decay of the charge density $\bar{\rho}(0,t)$ for an isolated positive charge initially placed at $\boldsymbol{x}=0$ in the bulk of the system: see a). The decay is consistent with diffusion in two dimensions. \textbf{c)} Scaling collapse of the charge distribution at different times along $\bar{\rho}(\boldsymbol{x}=(x,0),t)$, indicating Gaussian diffusion. Numerical results were averaged over $3\times 10^6$ runs of the random time evolution~\cite{sm}(Sec.\,G).}
\label{fig:dipole_dimer}
\end{figure}

To illustrate the generality of the condition \eqnref{eq:dipole5}, we consider systems beyond 1D. 
In analogy to $d$\,$=$\,$1$, for $d$\,$>$\,$1$ we write $\uc(\boldsymbol{x})$\,$=$\,$\boldsymbol{\nabla}\cdot \hat{\boldsymbol{e}}(\boldsymbol{x})$,
where $\hat{\boldsymbol{e}}(\boldsymbol{x})$\,$=$\,$(\hat{e}_1(\boldsymbol{x}),...,\hat{e}_d(\boldsymbol{x}))$ is now a $d$-component vector. We recognize that $\hat{\boldsymbol{e}}(\boldsymbol{x})$ is not uniquely determined by the charge configuration $\uc(\boldsymbol{x})$, and the relation between these variables takes the form of a $U(1)$ Gauss law, where the $\hat{\boldsymbol{e}}(\boldsymbol{x})$ constitute electric field degrees of freedom. Indeed, area-law charge fluctuations arise in $U(1)$ gauge theories if fluctuations of the electric fields $\hat{\boldsymbol{e}}(\boldsymbol{x})$ are bounded, as $\int_V dV \uc(\boldsymbol{x})$\,$=$\,$\int_{\partial V} d\boldsymbol{A}\cdot \hat{\boldsymbol{e}}(\boldsymbol{x})$. Thus, imposing global dipole conservation on $U(1)$ link models~\cite{horn1981finite,orland1990lattice,chandrasekharan1997quantum} with a finite electric field state space gives rise to diffusive behavior through \eqnref{eq:dipole5}.
To verify this prediction, we numerically simulate classical, discrete random time evolution in a hard-core dimer model on a square lattice (see \figref{fig:dipole_dimer}{a}), which can be mapped to a $U(1)$ link model~\cite{fradkin1990short,moessner_rvb}.
Under this mapping, a site $\boldsymbol{x}$ without any attached dimer carries a charge $\uc(\boldsymbol{x})$\,$=$\,$(-1)^{x_1+x_2}$ at $\boldsymbol{x}$\,$=$\,$(x_1,x_2)$, while a site with an attached dimer carries no charge. In the dynamics carried out numerically (see~\cite{sm} Sec.\,G, as well as Refs.~\cite{Iaconis19,morningstar2020kinetically,feldmeier2020anomalous,iaconis2021_multipole} for related approaches), we then explicitly incorporate conservation of the dipole moment associated to $\uc(\boldsymbol{x})$.
Starting from an initial state with an isolated positive charge in the bulk of the system $\uc(\boldsymbol{x},t$\,$=$\,$0)$\,$=$\,$\delta_{x_1,0}\, \delta_{x_2,0}$ (see \figref{fig:dipole_dimer}{a}), we numerically find a diffusive broadening of the resulting charge distribution at late times. As the overall charge density in the system vanishes, and positive and negative charges occupy different sublattices, we consider the quantity $\bar{\rho}(x_1,t)$\,$\equiv$\,$\uc((x_1,0),t)+\uc((x_1-1,0),t)$. We show in \figref{fig:dipole_dimer}{c} that $t\, \bar{\rho}(x_1,t)$ exhibits a scaling collapse when plotted against $x_1/\sqrt{t}$, in agreement with diffusive relaxation in two dimensions.


\emph{Conclusion and Outlook.} In this work we have established a comprehensive understanding of conserved operator dynamics under Brownian random unitary time evolution through a duality with the spectral properties of an associated effective Hamiltonian. Though the U(1) symmetric Brownian
evolution was used for clarity of presentation, these results generalize for any dynamics conserving a continuous global symmetry governed by a Lindblad equation~\cite{sm}(Sec.\,C). 
As the groundstate manifold always exhibits a spontaneous symmetry breaking of a continuous symmetry, a single-mode approximation could be applied to capture the low energy physics of this effective Hamiltonian to reproduce a number of dynamical universality classes for short- and long-range interacting systems with charge and multipole conservation laws. In addition, our formalism allowed us to study the Krylov-space-resolved hydrodynamics of dipole-conserving systems, establishing diffusive behavior in Krylov spaces with area law charge fluctuations, in contrast to more generic dynamics in the presence of dipole conservation.

We expect that such diffusive relaxation in dipole-conserving systems is valid beyond the specific examples studied numerically here and holds whenever the time evolution proceeds within an \textit{effective} state space (not necessarily a Krylov space) that fulfills \eqnref{eq:dipole5}. In particular, bounded fluctuations of the variables $\hat{e}_x$ can arise from energetics, for example via a term $\sim$\,$(\hat{e}_x)^2$ in the Hamiltonian, as appears naturally in standard electromagnetism. In this context, the resulting area-law charge fluctuations can be interpreted as Coulomb repulsion, which consequently leads to diffusive relaxation in dipole-conserving systems.
Furthermore, bounded charge fluctuations occur in many other interesting models: It was shown in Refs.\cite{lake2022dipole,zechmann2022fractonic} that area law charge fluctuations can arise in dipole-conserving Bose-Hubbard models in low-energy Mott states whenever a finite energy gap exists for charged excitations. It would be interesting to study the relevance of our results to such systems in the future.

\bigskip

\emph{Acknowledgements.} The authors thank Julian Bösl, Amos Chan, Soonwon Choi, Luca Delacretaz, Michael Knap, Ethan Lake, and Philip Zechmann for insightful discussions. 
O.O is supported by the Science and Technology Center for Integrated Quantum Materials, National Science Foundation Grant No. DMR1231319. 
J.F. acknowledges support by the Harvard Quantum Initiative.
J.Y.L is supported by the Gordon and Betty Moore Foundation under the grant GBMF8690 and by the National Science Foundation under the grant PHY-1748958.

\emph{Note added}: While finalizing this work, we became aware of independent related works~\cite{PhysRevB.108.L020304, PhysRevB.108.195106}. The previous version~\cite{prev} considered an unbounded single-mode dispersion for effectively all-to-all interacting models, leading to ultra-fast relaxation. Renormalizing the dispersion to be bounded results in a finite dynamical exponent for $\alpha_0\leq d/2$ as shown in \figref{fig:diffuse_phase}.

\bibliography{ref}

\begin{thebibliography}{83}%
\makeatletter
\providecommand \@ifxundefined [1]{%
 \@ifx{#1\undefined}
}%
\providecommand \@ifnum [1]{%
 \ifnum #1\expandafter \@firstoftwo
 \else \expandafter \@secondoftwo
 \fi
}%
\providecommand \@ifx [1]{%
 \ifx #1\expandafter \@firstoftwo
 \else \expandafter \@secondoftwo
 \fi
}%
\providecommand \natexlab [1]{#1}%
\providecommand \enquote  [1]{``#1''}%
\providecommand \bibnamefont  [1]{#1}%
\providecommand \bibfnamefont [1]{#1}%
\providecommand \citenamefont [1]{#1}%
\providecommand \href@noop [0]{\@secondoftwo}%
\providecommand \href [0]{\begingroup \@sanitize@url \@href}%
\providecommand \@href[1]{\@@startlink{#1}\@@href}%
\providecommand \@@href[1]{\endgroup#1\@@endlink}%
\providecommand \@sanitize@url [0]{\catcode `\\12\catcode `\$12\catcode
  `\&12\catcode `\#12\catcode `\^12\catcode `\_12\catcode `\%12\relax}%
\providecommand \@@startlink[1]{}%
\providecommand \@@endlink[0]{}%
\providecommand \url  [0]{\begingroup\@sanitize@url \@url }%
\providecommand \@url [1]{\endgroup\@href {#1}{\urlprefix }}%
\providecommand \urlprefix  [0]{URL }%
\providecommand \Eprint [0]{\href }%
\providecommand \doibase [0]{https://doi.org/}%
\providecommand \selectlanguage [0]{\@gobble}%
\providecommand \bibinfo  [0]{\@secondoftwo}%
\providecommand \bibfield  [0]{\@secondoftwo}%
\providecommand \translation [1]{[#1]}%
\providecommand \BibitemOpen [0]{}%
\providecommand \bibitemStop [0]{}%
\providecommand \bibitemNoStop [0]{.\EOS\space}%
\providecommand \EOS [0]{\spacefactor3000\relax}%
\providecommand \BibitemShut  [1]{\csname bibitem#1\endcsname}%
\let\auto@bib@innerbib\@empty
\bibitem [{\citenamefont {Deutsch}(1991)}]{Deutsch1991}%
  \BibitemOpen
  \bibfield  {author} {\bibinfo {author} {\bibfnamefont {J.~M.}\ \bibnamefont
  {Deutsch}},\ }\bibfield  {title} {\bibinfo {title} {Quantum statistical
  mechanics in a closed system},\ }\href
  {https://doi.org/10.1103/PhysRevA.43.2046} {\bibfield  {journal} {\bibinfo
  {journal} {Phys. Rev. A}\ }\textbf {\bibinfo {volume} {43}},\ \bibinfo
  {pages} {2046} (\bibinfo {year} {1991})}\BibitemShut {NoStop}%
\bibitem [{\citenamefont {Srednicki}(1994)}]{Srednicki1994}%
  \BibitemOpen
  \bibfield  {author} {\bibinfo {author} {\bibfnamefont {M.}~\bibnamefont
  {Srednicki}},\ }\bibfield  {title} {\bibinfo {title} {Chaos and quantum
  thermalization},\ }\href {https://doi.org/10.1103/PhysRevE.50.888} {\bibfield
   {journal} {\bibinfo  {journal} {Phys. Rev. E}\ }\textbf {\bibinfo {volume}
  {50}},\ \bibinfo {pages} {888} (\bibinfo {year} {1994})}\BibitemShut
  {NoStop}%
\bibitem [{\citenamefont {Rigol}\ \emph {et~al.}(2008)\citenamefont {Rigol},
  \citenamefont {Dunjko},\ and\ \citenamefont {Olshanii}}]{Rigol2008}%
  \BibitemOpen
  \bibfield  {author} {\bibinfo {author} {\bibfnamefont {M.}~\bibnamefont
  {Rigol}}, \bibinfo {author} {\bibfnamefont {V.}~\bibnamefont {Dunjko}},\ and\
  \bibinfo {author} {\bibfnamefont {M.}~\bibnamefont {Olshanii}},\ }\bibfield
  {title} {\bibinfo {title} {Thermalization and its mechanism for generic
  isolated quantum systems},\ }\href {https://doi.org/10.1038/nature06838}
  {\bibfield  {journal} {\bibinfo  {journal} {Nature}\ }\textbf {\bibinfo
  {volume} {452}},\ \bibinfo {pages} {854} (\bibinfo {year}
  {2008})}\BibitemShut {NoStop}%
\bibitem [{\citenamefont {Polkovnikov}\ \emph {et~al.}(2011)\citenamefont
  {Polkovnikov}, \citenamefont {Sengupta}, \citenamefont {Silva},\ and\
  \citenamefont {Vengalattore}}]{Polkovnikov2011}%
  \BibitemOpen
  \bibfield  {author} {\bibinfo {author} {\bibfnamefont {A.}~\bibnamefont
  {Polkovnikov}}, \bibinfo {author} {\bibfnamefont {K.}~\bibnamefont
  {Sengupta}}, \bibinfo {author} {\bibfnamefont {A.}~\bibnamefont {Silva}},\
  and\ \bibinfo {author} {\bibfnamefont {M.}~\bibnamefont {Vengalattore}},\
  }\bibfield  {title} {\bibinfo {title} {Colloquium: Nonequilibrium dynamics of
  closed interacting quantum systems},\ }\href
  {https://doi.org/10.1103/RevModPhys.83.863} {\bibfield  {journal} {\bibinfo
  {journal} {Rev. Mod. Phys.}\ }\textbf {\bibinfo {volume} {83}},\ \bibinfo
  {pages} {863} (\bibinfo {year} {2011})}\BibitemShut {NoStop}%
\bibitem [{\citenamefont {D'Alessio}\ \emph {et~al.}(2016)\citenamefont
  {D'Alessio}, \citenamefont {Kafri}, \citenamefont {Polkovnikov},\ and\
  \citenamefont {Rigol}}]{Alessio2016}%
  \BibitemOpen
  \bibfield  {author} {\bibinfo {author} {\bibfnamefont {L.}~\bibnamefont
  {D'Alessio}}, \bibinfo {author} {\bibfnamefont {Y.}~\bibnamefont {Kafri}},
  \bibinfo {author} {\bibfnamefont {A.}~\bibnamefont {Polkovnikov}},\ and\
  \bibinfo {author} {\bibfnamefont {M.}~\bibnamefont {Rigol}},\ }\bibfield
  {title} {\bibinfo {title} {From quantum chaos and eigenstate thermalization
  to statistical mechanics and thermodynamics},\ }\href
  {https://doi.org/10.1080/00018732.2016.1198134} {\bibfield  {journal}
  {\bibinfo  {journal} {Advances in Physics}\ }\textbf {\bibinfo {volume}
  {65}},\ \bibinfo {pages} {239} (\bibinfo {year} {2016})},\ \Eprint
  {https://arxiv.org/abs/https://doi.org/10.1080/00018732.2016.1198134}
  {https://doi.org/10.1080/00018732.2016.1198134} \BibitemShut {NoStop}%
\bibitem [{\citenamefont {Kaufman}\ \emph {et~al.}(2016)\citenamefont
  {Kaufman}, \citenamefont {Tai}, \citenamefont {Lukin}, \citenamefont
  {Rispoli}, \citenamefont {Schittko}, \citenamefont {Preiss},\ and\
  \citenamefont {Greiner}}]{kaufman2016_thermalization}%
  \BibitemOpen
  \bibfield  {author} {\bibinfo {author} {\bibfnamefont {A.~M.}\ \bibnamefont
  {Kaufman}}, \bibinfo {author} {\bibfnamefont {M.~E.}\ \bibnamefont {Tai}},
  \bibinfo {author} {\bibfnamefont {A.}~\bibnamefont {Lukin}}, \bibinfo
  {author} {\bibfnamefont {M.}~\bibnamefont {Rispoli}}, \bibinfo {author}
  {\bibfnamefont {R.}~\bibnamefont {Schittko}}, \bibinfo {author}
  {\bibfnamefont {P.~M.}\ \bibnamefont {Preiss}},\ and\ \bibinfo {author}
  {\bibfnamefont {M.}~\bibnamefont {Greiner}},\ }\bibfield  {title} {\bibinfo
  {title} {{Quantum thermalization through entanglement in an isolated
  many-body system}},\ }\href {https://doi.org/10.1126/science.aaf6725}
  {\bibfield  {journal} {\bibinfo  {journal} {Science}\ }\textbf {\bibinfo
  {volume} {353}},\ \bibinfo {pages} {794} (\bibinfo {year}
  {2016})}\BibitemShut {NoStop}%
\bibitem [{\citenamefont {Brydges}\ \emph {et~al.}(2019)\citenamefont
  {Brydges}, \citenamefont {Elben}, \citenamefont {Jurcevic}, \citenamefont
  {Vermersch}, \citenamefont {Maier}, \citenamefont {Lanyon}, \citenamefont
  {Zoller}, \citenamefont {Blatt},\ and\ \citenamefont
  {Roos}}]{Brydges2019_renyi}%
  \BibitemOpen
  \bibfield  {author} {\bibinfo {author} {\bibfnamefont {T.}~\bibnamefont
  {Brydges}}, \bibinfo {author} {\bibfnamefont {A.}~\bibnamefont {Elben}},
  \bibinfo {author} {\bibfnamefont {P.}~\bibnamefont {Jurcevic}}, \bibinfo
  {author} {\bibfnamefont {B.}~\bibnamefont {Vermersch}}, \bibinfo {author}
  {\bibfnamefont {C.}~\bibnamefont {Maier}}, \bibinfo {author} {\bibfnamefont
  {B.~P.}\ \bibnamefont {Lanyon}}, \bibinfo {author} {\bibfnamefont
  {P.}~\bibnamefont {Zoller}}, \bibinfo {author} {\bibfnamefont
  {R.}~\bibnamefont {Blatt}},\ and\ \bibinfo {author} {\bibfnamefont {C.~F.}\
  \bibnamefont {Roos}},\ }\bibfield  {title} {\bibinfo {title} {{Probing
  R{\'e}nyi entanglement entropy via randomized measurements}},\ }\href
  {https://doi.org/10.1126/science.aau4963} {\bibfield  {journal} {\bibinfo
  {journal} {Science}\ }\textbf {\bibinfo {volume} {364}},\ \bibinfo {pages}
  {260} (\bibinfo {year} {2019})}\BibitemShut {NoStop}%
\bibitem [{\citenamefont {Nahum}\ \emph {et~al.}(2017)\citenamefont {Nahum},
  \citenamefont {Ruhman}, \citenamefont {Vijay},\ and\ \citenamefont
  {Haah}}]{Nahum2017}%
  \BibitemOpen
  \bibfield  {author} {\bibinfo {author} {\bibfnamefont {A.}~\bibnamefont
  {Nahum}}, \bibinfo {author} {\bibfnamefont {J.}~\bibnamefont {Ruhman}},
  \bibinfo {author} {\bibfnamefont {S.}~\bibnamefont {Vijay}},\ and\ \bibinfo
  {author} {\bibfnamefont {J.}~\bibnamefont {Haah}},\ }\bibfield  {title}
  {\bibinfo {title} {Quantum entanglement growth under random unitary
  dynamics},\ }\href {https://doi.org/10.1103/PhysRevX.7.031016} {\bibfield
  {journal} {\bibinfo  {journal} {Phys. Rev. X}\ }\textbf {\bibinfo {volume}
  {7}},\ \bibinfo {pages} {031016} (\bibinfo {year} {2017})}\BibitemShut
  {NoStop}%
\bibitem [{\citenamefont {von Keyserlingk}\ \emph {et~al.}(2018)\citenamefont
  {von Keyserlingk}, \citenamefont {Rakovszky}, \citenamefont {Pollmann},\ and\
  \citenamefont {Sondhi}}]{Keyserlingk2018}%
  \BibitemOpen
  \bibfield  {author} {\bibinfo {author} {\bibfnamefont {C.~W.}\ \bibnamefont
  {von Keyserlingk}}, \bibinfo {author} {\bibfnamefont {T.}~\bibnamefont
  {Rakovszky}}, \bibinfo {author} {\bibfnamefont {F.}~\bibnamefont
  {Pollmann}},\ and\ \bibinfo {author} {\bibfnamefont {S.~L.}\ \bibnamefont
  {Sondhi}},\ }\bibfield  {title} {\bibinfo {title} {Operator hydrodynamics,
  otocs, and entanglement growth in systems without conservation laws},\ }\href
  {https://doi.org/10.1103/PhysRevX.8.021013} {\bibfield  {journal} {\bibinfo
  {journal} {Phys. Rev. X}\ }\textbf {\bibinfo {volume} {8}},\ \bibinfo {pages}
  {021013} (\bibinfo {year} {2018})}\BibitemShut {NoStop}%
\bibitem [{\citenamefont {Hamma}\ \emph
  {et~al.}(2012{\natexlab{a}})\citenamefont {Hamma}, \citenamefont {Santra},\
  and\ \citenamefont {Zanardi}}]{Hamma2012}%
  \BibitemOpen
  \bibfield  {author} {\bibinfo {author} {\bibfnamefont {A.}~\bibnamefont
  {Hamma}}, \bibinfo {author} {\bibfnamefont {S.}~\bibnamefont {Santra}},\ and\
  \bibinfo {author} {\bibfnamefont {P.}~\bibnamefont {Zanardi}},\ }\bibfield
  {title} {\bibinfo {title} {Quantum entanglement in random physical states},\
  }\href {https://doi.org/10.1103/PhysRevLett.109.040502} {\bibfield  {journal}
  {\bibinfo  {journal} {Phys. Rev. Lett.}\ }\textbf {\bibinfo {volume} {109}},\
  \bibinfo {pages} {040502} (\bibinfo {year} {2012}{\natexlab{a}})}\BibitemShut
  {NoStop}%
\bibitem [{\citenamefont {Hamma}\ \emph
  {et~al.}(2012{\natexlab{b}})\citenamefont {Hamma}, \citenamefont {Santra},\
  and\ \citenamefont {Zanardi}}]{Hamma2012b}%
  \BibitemOpen
  \bibfield  {author} {\bibinfo {author} {\bibfnamefont {A.}~\bibnamefont
  {Hamma}}, \bibinfo {author} {\bibfnamefont {S.}~\bibnamefont {Santra}},\ and\
  \bibinfo {author} {\bibfnamefont {P.}~\bibnamefont {Zanardi}},\ }\bibfield
  {title} {\bibinfo {title} {Ensembles of physical states and random quantum
  circuits on graphs},\ }\href {https://doi.org/10.1103/PhysRevA.86.052324}
  {\bibfield  {journal} {\bibinfo  {journal} {Phys. Rev. A}\ }\textbf {\bibinfo
  {volume} {86}},\ \bibinfo {pages} {052324} (\bibinfo {year}
  {2012}{\natexlab{b}})}\BibitemShut {NoStop}%
\bibitem [{\citenamefont {Kos}\ \emph {et~al.}(2018)\citenamefont {Kos},
  \citenamefont {Ljubotina},\ and\ \citenamefont {Prosen}}]{Pavel2018}%
  \BibitemOpen
  \bibfield  {author} {\bibinfo {author} {\bibfnamefont {P.}~\bibnamefont
  {Kos}}, \bibinfo {author} {\bibfnamefont {M.}~\bibnamefont {Ljubotina}},\
  and\ \bibinfo {author} {\bibfnamefont {T.~c.~v.}\ \bibnamefont {Prosen}},\
  }\bibfield  {title} {\bibinfo {title} {Many-body quantum chaos: Analytic
  connection to random matrix theory},\ }\href
  {https://doi.org/10.1103/PhysRevX.8.021062} {\bibfield  {journal} {\bibinfo
  {journal} {Phys. Rev. X}\ }\textbf {\bibinfo {volume} {8}},\ \bibinfo {pages}
  {021062} (\bibinfo {year} {2018})}\BibitemShut {NoStop}%
\bibitem [{\citenamefont {Gharibyan}\ \emph {et~al.}(2018)\citenamefont
  {Gharibyan}, \citenamefont {Hanada}, \citenamefont {Shenker},\ and\
  \citenamefont {Tezuka}}]{Gharibyan2018}%
  \BibitemOpen
  \bibfield  {author} {\bibinfo {author} {\bibfnamefont {H.}~\bibnamefont
  {Gharibyan}}, \bibinfo {author} {\bibfnamefont {M.}~\bibnamefont {Hanada}},
  \bibinfo {author} {\bibfnamefont {S.~H.}\ \bibnamefont {Shenker}},\ and\
  \bibinfo {author} {\bibfnamefont {M.}~\bibnamefont {Tezuka}},\ }\bibfield
  {title} {\bibinfo {title} {Onset of random matrix behavior in scrambling
  systems},\ }\href {https://doi.org/10.1007/JHEP07(2018)124} {\bibfield
  {journal} {\bibinfo  {journal} {Journal of High Energy Physics}\ }\textbf
  {\bibinfo {volume} {2018}},\ \bibinfo {pages} {124} (\bibinfo {year}
  {2018})}\BibitemShut {NoStop}%
\bibitem [{\citenamefont {Moudgalya}\ \emph {et~al.}(2019)\citenamefont
  {Moudgalya}, \citenamefont {Devakul}, \citenamefont {von Keyserlingk},\ and\
  \citenamefont {Sondhi}}]{Moudgalya2019}%
  \BibitemOpen
  \bibfield  {author} {\bibinfo {author} {\bibfnamefont {S.}~\bibnamefont
  {Moudgalya}}, \bibinfo {author} {\bibfnamefont {T.}~\bibnamefont {Devakul}},
  \bibinfo {author} {\bibfnamefont {C.~W.}\ \bibnamefont {von Keyserlingk}},\
  and\ \bibinfo {author} {\bibfnamefont {S.~L.}\ \bibnamefont {Sondhi}},\
  }\bibfield  {title} {\bibinfo {title} {Operator spreading in quantum maps},\
  }\href {https://doi.org/10.1103/PhysRevB.99.094312} {\bibfield  {journal}
  {\bibinfo  {journal} {Phys. Rev. B}\ }\textbf {\bibinfo {volume} {99}},\
  \bibinfo {pages} {094312} (\bibinfo {year} {2019})}\BibitemShut {NoStop}%
\bibitem [{\citenamefont {Khemani}\ \emph {et~al.}(2018)\citenamefont
  {Khemani}, \citenamefont {Vishwanath},\ and\ \citenamefont
  {Huse}}]{Vedika2018}%
  \BibitemOpen
  \bibfield  {author} {\bibinfo {author} {\bibfnamefont {V.}~\bibnamefont
  {Khemani}}, \bibinfo {author} {\bibfnamefont {A.}~\bibnamefont
  {Vishwanath}},\ and\ \bibinfo {author} {\bibfnamefont {D.~A.}\ \bibnamefont
  {Huse}},\ }\bibfield  {title} {\bibinfo {title} {Operator spreading and the
  emergence of dissipative hydrodynamics under unitary evolution with
  conservation laws},\ }\href {https://doi.org/10.1103/PhysRevX.8.031057}
  {\bibfield  {journal} {\bibinfo  {journal} {Phys. Rev. X}\ }\textbf {\bibinfo
  {volume} {8}},\ \bibinfo {pages} {031057} (\bibinfo {year}
  {2018})}\BibitemShut {NoStop}%
\bibitem [{\citenamefont {Rakovszky}\ \emph {et~al.}(2018)\citenamefont
  {Rakovszky}, \citenamefont {Pollmann},\ and\ \citenamefont {von
  Keyserlingk}}]{Rakovszky2018}%
  \BibitemOpen
  \bibfield  {author} {\bibinfo {author} {\bibfnamefont {T.}~\bibnamefont
  {Rakovszky}}, \bibinfo {author} {\bibfnamefont {F.}~\bibnamefont
  {Pollmann}},\ and\ \bibinfo {author} {\bibfnamefont {C.~W.}\ \bibnamefont
  {von Keyserlingk}},\ }\bibfield  {title} {\bibinfo {title} {Diffusive
  hydrodynamics of out-of-time-ordered correlators with charge conservation},\
  }\href {https://doi.org/10.1103/PhysRevX.8.031058} {\bibfield  {journal}
  {\bibinfo  {journal} {Phys. Rev. X}\ }\textbf {\bibinfo {volume} {8}},\
  \bibinfo {pages} {031058} (\bibinfo {year} {2018})}\BibitemShut {NoStop}%
\bibitem [{\citenamefont {Schuckert}\ \emph {et~al.}(2020)\citenamefont
  {Schuckert}, \citenamefont {Lovas},\ and\ \citenamefont
  {Knap}}]{Schuckert2020LR}%
  \BibitemOpen
  \bibfield  {author} {\bibinfo {author} {\bibfnamefont {A.}~\bibnamefont
  {Schuckert}}, \bibinfo {author} {\bibfnamefont {I.}~\bibnamefont {Lovas}},\
  and\ \bibinfo {author} {\bibfnamefont {M.}~\bibnamefont {Knap}},\ }\bibfield
  {title} {\bibinfo {title} {Nonlocal emergent hydrodynamics in a long-range
  quantum spin system},\ }\href {https://doi.org/10.1103/PhysRevB.101.020416}
  {\bibfield  {journal} {\bibinfo  {journal} {Phys. Rev. B}\ }\textbf {\bibinfo
  {volume} {101}},\ \bibinfo {pages} {020416(R)} (\bibinfo {year}
  {2020})}\BibitemShut {NoStop}%
\bibitem [{\citenamefont {Joshi}\ \emph {et~al.}(2022)\citenamefont {Joshi},
  \citenamefont {Kranzl}, \citenamefont {Schuckert}, \citenamefont {Lovas},
  \citenamefont {Maier}, \citenamefont {Blatt}, \citenamefont {Knap},\ and\
  \citenamefont {Roos}}]{Joshi2022LR}%
  \BibitemOpen
  \bibfield  {author} {\bibinfo {author} {\bibfnamefont {M.~K.}\ \bibnamefont
  {Joshi}}, \bibinfo {author} {\bibfnamefont {F.}~\bibnamefont {Kranzl}},
  \bibinfo {author} {\bibfnamefont {A.}~\bibnamefont {Schuckert}}, \bibinfo
  {author} {\bibfnamefont {I.}~\bibnamefont {Lovas}}, \bibinfo {author}
  {\bibfnamefont {C.}~\bibnamefont {Maier}}, \bibinfo {author} {\bibfnamefont
  {R.}~\bibnamefont {Blatt}}, \bibinfo {author} {\bibfnamefont
  {M.}~\bibnamefont {Knap}},\ and\ \bibinfo {author} {\bibfnamefont {C.~F.}\
  \bibnamefont {Roos}},\ }\bibfield  {title} {\bibinfo {title} {Observing
  emergent hydrodynamics in a long-range quantum magnet},\ }\href
  {https://doi.org/10.1126/science.abk2400} {\bibfield  {journal} {\bibinfo
  {journal} {Science}\ }\textbf {\bibinfo {volume} {376}},\ \bibinfo {pages}
  {720} (\bibinfo {year} {2022})}\BibitemShut {NoStop}%
\bibitem [{\citenamefont {Richter}\ \emph {et~al.}(2023)\citenamefont
  {Richter}, \citenamefont {Lunt},\ and\ \citenamefont
  {Pal}}]{richter2023_long}%
  \BibitemOpen
  \bibfield  {author} {\bibinfo {author} {\bibfnamefont {J.}~\bibnamefont
  {Richter}}, \bibinfo {author} {\bibfnamefont {O.}~\bibnamefont {Lunt}},\ and\
  \bibinfo {author} {\bibfnamefont {A.}~\bibnamefont {Pal}},\ }\bibfield
  {title} {\bibinfo {title} {Transport and entanglement growth in long-range
  random clifford circuits},\ }\href
  {https://doi.org/10.1103/PhysRevResearch.5.L012031} {\bibfield  {journal}
  {\bibinfo  {journal} {Phys. Rev. Res.}\ }\textbf {\bibinfo {volume} {5}},\
  \bibinfo {pages} {L012031} (\bibinfo {year} {2023})}\BibitemShut {NoStop}%
\bibitem [{\citenamefont {Zu}\ \emph {et~al.}(2021)\citenamefont {Zu},
  \citenamefont {Machado}, \citenamefont {Ye}, \citenamefont {Choi},
  \citenamefont {Kobrin}, \citenamefont {Mittiga}, \citenamefont {Hsieh},
  \citenamefont {Bhattacharyya}, \citenamefont {Markham}, \citenamefont
  {Twitchen} \emph {et~al.}}]{zu2021emergent}%
  \BibitemOpen
  \bibfield  {author} {\bibinfo {author} {\bibfnamefont {C.}~\bibnamefont
  {Zu}}, \bibinfo {author} {\bibfnamefont {F.}~\bibnamefont {Machado}},
  \bibinfo {author} {\bibfnamefont {B.}~\bibnamefont {Ye}}, \bibinfo {author}
  {\bibfnamefont {S.}~\bibnamefont {Choi}}, \bibinfo {author} {\bibfnamefont
  {B.}~\bibnamefont {Kobrin}}, \bibinfo {author} {\bibfnamefont
  {T.}~\bibnamefont {Mittiga}}, \bibinfo {author} {\bibfnamefont
  {S.}~\bibnamefont {Hsieh}}, \bibinfo {author} {\bibfnamefont
  {P.}~\bibnamefont {Bhattacharyya}}, \bibinfo {author} {\bibfnamefont
  {M.}~\bibnamefont {Markham}}, \bibinfo {author} {\bibfnamefont
  {D.}~\bibnamefont {Twitchen}}, \emph {et~al.},\ }\bibfield  {title} {\bibinfo
  {title} {Emergent hydrodynamics in a strongly interacting dipolar spin
  ensemble},\ }\href@noop {} {\bibfield  {journal} {\bibinfo  {journal}
  {Nature}\ }\textbf {\bibinfo {volume} {597}},\ \bibinfo {pages} {45}
  (\bibinfo {year} {2021})}\BibitemShut {NoStop}%
\bibitem [{\citenamefont {Iaconis}\ \emph {et~al.}(2019)\citenamefont
  {Iaconis}, \citenamefont {Vijay},\ and\ \citenamefont
  {Nandkishore}}]{Iaconis19}%
  \BibitemOpen
  \bibfield  {author} {\bibinfo {author} {\bibfnamefont {J.}~\bibnamefont
  {Iaconis}}, \bibinfo {author} {\bibfnamefont {S.}~\bibnamefont {Vijay}},\
  and\ \bibinfo {author} {\bibfnamefont {R.}~\bibnamefont {Nandkishore}},\
  }\bibfield  {title} {\bibinfo {title} {{Anomalous subdiffusion from subsystem
  symmetries}},\ }\href {https://doi.org/10.1103/PhysRevB.100.214301}
  {\bibfield  {journal} {\bibinfo  {journal} {Phys. Rev. B}\ }\textbf {\bibinfo
  {volume} {100}},\ \bibinfo {pages} {214301} (\bibinfo {year}
  {2019})}\BibitemShut {NoStop}%
\bibitem [{\citenamefont {Guardado-Sanchez}\ \emph {et~al.}(2020)\citenamefont
  {Guardado-Sanchez}, \citenamefont {Morningstar}, \citenamefont {Spar},
  \citenamefont {Brown}, \citenamefont {Huse},\ and\ \citenamefont
  {Bakr}}]{guardado2020subdiffusion}%
  \BibitemOpen
  \bibfield  {author} {\bibinfo {author} {\bibfnamefont {E.}~\bibnamefont
  {Guardado-Sanchez}}, \bibinfo {author} {\bibfnamefont {A.}~\bibnamefont
  {Morningstar}}, \bibinfo {author} {\bibfnamefont {B.~M.}\ \bibnamefont
  {Spar}}, \bibinfo {author} {\bibfnamefont {P.~T.}\ \bibnamefont {Brown}},
  \bibinfo {author} {\bibfnamefont {D.~A.}\ \bibnamefont {Huse}},\ and\
  \bibinfo {author} {\bibfnamefont {W.~S.}\ \bibnamefont {Bakr}},\ }\bibfield
  {title} {\bibinfo {title} {{Subdiffusion and heat transport in a tilted
  two-dimensional Fermi-Hubbard system}},\ }\href@noop {} {\bibfield  {journal}
  {\bibinfo  {journal} {Physical Review X}\ }\textbf {\bibinfo {volume} {10}},\
  \bibinfo {pages} {011042} (\bibinfo {year} {2020})}\BibitemShut {NoStop}%
\bibitem [{\citenamefont {Gromov}\ \emph {et~al.}(2020)\citenamefont {Gromov},
  \citenamefont {Lucas},\ and\ \citenamefont
  {Nandkishore}}]{gromov2020fracton}%
  \BibitemOpen
  \bibfield  {author} {\bibinfo {author} {\bibfnamefont {A.}~\bibnamefont
  {Gromov}}, \bibinfo {author} {\bibfnamefont {A.}~\bibnamefont {Lucas}},\ and\
  \bibinfo {author} {\bibfnamefont {R.~M.}\ \bibnamefont {Nandkishore}},\
  }\bibfield  {title} {\bibinfo {title} {Fracton hydrodynamics},\ }\href@noop
  {} {\bibfield  {journal} {\bibinfo  {journal} {Physical Review Research}\
  }\textbf {\bibinfo {volume} {2}},\ \bibinfo {pages} {033124} (\bibinfo {year}
  {2020})}\BibitemShut {NoStop}%
\bibitem [{\citenamefont {Feldmeier}\ \emph {et~al.}(2020)\citenamefont
  {Feldmeier}, \citenamefont {Sala}, \citenamefont {De~Tomasi}, \citenamefont
  {Pollmann},\ and\ \citenamefont {Knap}}]{feldmeier2020anomalous}%
  \BibitemOpen
  \bibfield  {author} {\bibinfo {author} {\bibfnamefont {J.}~\bibnamefont
  {Feldmeier}}, \bibinfo {author} {\bibfnamefont {P.}~\bibnamefont {Sala}},
  \bibinfo {author} {\bibfnamefont {G.}~\bibnamefont {De~Tomasi}}, \bibinfo
  {author} {\bibfnamefont {F.}~\bibnamefont {Pollmann}},\ and\ \bibinfo
  {author} {\bibfnamefont {M.}~\bibnamefont {Knap}},\ }\bibfield  {title}
  {\bibinfo {title} {Anomalous diffusion in dipole-and higher-moment-conserving
  systems},\ }\href@noop {} {\bibfield  {journal} {\bibinfo  {journal}
  {Physical Review Letters}\ }\textbf {\bibinfo {volume} {125}},\ \bibinfo
  {pages} {245303} (\bibinfo {year} {2020})}\BibitemShut {NoStop}%
\bibitem [{\citenamefont {Morningstar}\ \emph {et~al.}(2020)\citenamefont
  {Morningstar}, \citenamefont {Khemani},\ and\ \citenamefont
  {Huse}}]{morningstar2020kinetically}%
  \BibitemOpen
  \bibfield  {author} {\bibinfo {author} {\bibfnamefont {A.}~\bibnamefont
  {Morningstar}}, \bibinfo {author} {\bibfnamefont {V.}~\bibnamefont
  {Khemani}},\ and\ \bibinfo {author} {\bibfnamefont {D.~A.}\ \bibnamefont
  {Huse}},\ }\bibfield  {title} {\bibinfo {title} {Kinetically constrained
  freezing transition in a dipole-conserving system},\ }\href@noop {}
  {\bibfield  {journal} {\bibinfo  {journal} {Physical Review B}\ }\textbf
  {\bibinfo {volume} {101}},\ \bibinfo {pages} {214205} (\bibinfo {year}
  {2020})}\BibitemShut {NoStop}%
\bibitem [{\citenamefont {Zhang}(2020)}]{zhang_2020}%
  \BibitemOpen
  \bibfield  {author} {\bibinfo {author} {\bibfnamefont {P.}~\bibnamefont
  {Zhang}},\ }\bibfield  {title} {\bibinfo {title} {Subdiffusion in strongly
  tilted lattice systems},\ }\href
  {https://doi.org/10.1103/PhysRevResearch.2.033129} {\bibfield  {journal}
  {\bibinfo  {journal} {Phys. Rev. Research}\ }\textbf {\bibinfo {volume}
  {2}},\ \bibinfo {pages} {033129} (\bibinfo {year} {2020})}\BibitemShut
  {NoStop}%
\bibitem [{\citenamefont {Moudgalya}\ \emph {et~al.}(2021)\citenamefont
  {Moudgalya}, \citenamefont {Prem}, \citenamefont {Huse},\ and\ \citenamefont
  {Chan}}]{Moudgalya2021}%
  \BibitemOpen
  \bibfield  {author} {\bibinfo {author} {\bibfnamefont {S.}~\bibnamefont
  {Moudgalya}}, \bibinfo {author} {\bibfnamefont {A.}~\bibnamefont {Prem}},
  \bibinfo {author} {\bibfnamefont {D.~A.}\ \bibnamefont {Huse}},\ and\
  \bibinfo {author} {\bibfnamefont {A.}~\bibnamefont {Chan}},\ }\bibfield
  {title} {\bibinfo {title} {Spectral statistics in constrained many-body
  quantum chaotic systems},\ }\href
  {https://doi.org/10.1103/PhysRevResearch.3.023176} {\bibfield  {journal}
  {\bibinfo  {journal} {Phys. Rev. Res.}\ }\textbf {\bibinfo {volume} {3}},\
  \bibinfo {pages} {023176} (\bibinfo {year} {2021})}\BibitemShut {NoStop}%
\bibitem [{\citenamefont {{Singh, H. and Ware, B. A. and Vasseur, R. and
  Friedman, A. J.}}(2021)}]{Singh2021_subdiff}%
  \BibitemOpen
  \bibfield  {author} {\bibinfo {author} {\bibnamefont {{Singh, H. and Ware, B.
  A. and Vasseur, R. and Friedman, A. J.}}},\ }\bibfield  {title} {\bibinfo
  {title} {{Subdiffusion and Many-Body Quantum Chaos with Kinetic
  Constraints}},\ }\href {https://doi.org/10.1103/PhysRevLett.127.230602}
  {\bibfield  {journal} {\bibinfo  {journal} {Phys. Rev. Lett.}\ }\textbf
  {\bibinfo {volume} {127}},\ \bibinfo {pages} {230602} (\bibinfo {year}
  {2021})}\BibitemShut {NoStop}%
\bibitem [{\citenamefont {Feldmeier}\ \emph {et~al.}(2021)\citenamefont
  {Feldmeier}, \citenamefont {Pollmann},\ and\ \citenamefont
  {Knap}}]{feldmeier2021_fractondimer}%
  \BibitemOpen
  \bibfield  {author} {\bibinfo {author} {\bibfnamefont {J.}~\bibnamefont
  {Feldmeier}}, \bibinfo {author} {\bibfnamefont {F.}~\bibnamefont
  {Pollmann}},\ and\ \bibinfo {author} {\bibfnamefont {M.}~\bibnamefont
  {Knap}},\ }\bibfield  {title} {\bibinfo {title} {{Emergent fracton dynamics
  in a nonplanar dimer model}},\ }\href
  {https://doi.org/10.1103/PhysRevB.103.094303} {\bibfield  {journal} {\bibinfo
   {journal} {Phys. Rev. B}\ }\textbf {\bibinfo {volume} {103}},\ \bibinfo
  {pages} {094303} (\bibinfo {year} {2021})}\BibitemShut {NoStop}%
\bibitem [{\citenamefont {Sala}\ \emph {et~al.}(2022)\citenamefont {Sala},
  \citenamefont {Lehmann}, \citenamefont {Rakovszky},\ and\ \citenamefont
  {Pollmann}}]{sala2022_modulated}%
  \BibitemOpen
  \bibfield  {author} {\bibinfo {author} {\bibfnamefont {P.}~\bibnamefont
  {Sala}}, \bibinfo {author} {\bibfnamefont {J.}~\bibnamefont {Lehmann}},
  \bibinfo {author} {\bibfnamefont {T.}~\bibnamefont {Rakovszky}},\ and\
  \bibinfo {author} {\bibfnamefont {F.}~\bibnamefont {Pollmann}},\ }\bibfield
  {title} {\bibinfo {title} {Dynamics in systems with modulated symmetries},\
  }\href {https://doi.org/10.1103/PhysRevLett.129.170601} {\bibfield  {journal}
  {\bibinfo  {journal} {Phys. Rev. Lett.}\ }\textbf {\bibinfo {volume} {129}},\
  \bibinfo {pages} {170601} (\bibinfo {year} {2022})}\BibitemShut {NoStop}%
\bibitem [{\citenamefont {Hart}\ \emph {et~al.}(2022)\citenamefont {Hart},
  \citenamefont {Lucas},\ and\ \citenamefont {Nandkishore}}]{hart2022_hidden}%
  \BibitemOpen
  \bibfield  {author} {\bibinfo {author} {\bibfnamefont {O.}~\bibnamefont
  {Hart}}, \bibinfo {author} {\bibfnamefont {A.}~\bibnamefont {Lucas}},\ and\
  \bibinfo {author} {\bibfnamefont {R.}~\bibnamefont {Nandkishore}},\
  }\bibfield  {title} {\bibinfo {title} {Hidden quasiconservation laws in
  fracton hydrodynamics},\ }\href {https://doi.org/10.1103/PhysRevE.105.044103}
  {\bibfield  {journal} {\bibinfo  {journal} {Phys. Rev. E}\ }\textbf {\bibinfo
  {volume} {105}},\ \bibinfo {pages} {044103} (\bibinfo {year}
  {2022})}\BibitemShut {NoStop}%
\bibitem [{\citenamefont {Feldmeier}\ \emph {et~al.}(2022)\citenamefont
  {Feldmeier}, \citenamefont {Witczak-Krempa},\ and\ \citenamefont
  {Knap}}]{feldmeier2022_tracer}%
  \BibitemOpen
  \bibfield  {author} {\bibinfo {author} {\bibfnamefont {J.}~\bibnamefont
  {Feldmeier}}, \bibinfo {author} {\bibfnamefont {W.}~\bibnamefont
  {Witczak-Krempa}},\ and\ \bibinfo {author} {\bibfnamefont {M.}~\bibnamefont
  {Knap}},\ }\bibfield  {title} {\bibinfo {title} {Emergent tracer dynamics in
  constrained quantum systems},\ }\href
  {https://doi.org/10.1103/PhysRevB.106.094303} {\bibfield  {journal} {\bibinfo
   {journal} {Phys. Rev. B}\ }\textbf {\bibinfo {volume} {106}},\ \bibinfo
  {pages} {094303} (\bibinfo {year} {2022})}\BibitemShut {NoStop}%
\bibitem [{\citenamefont {Sala}\ \emph {et~al.}(2020)\citenamefont {Sala},
  \citenamefont {Rakovszky}, \citenamefont {Verresen}, \citenamefont {Knap},\
  and\ \citenamefont {Pollmann}}]{Sala2020}%
  \BibitemOpen
  \bibfield  {author} {\bibinfo {author} {\bibfnamefont {P.}~\bibnamefont
  {Sala}}, \bibinfo {author} {\bibfnamefont {T.}~\bibnamefont {Rakovszky}},
  \bibinfo {author} {\bibfnamefont {R.}~\bibnamefont {Verresen}}, \bibinfo
  {author} {\bibfnamefont {M.}~\bibnamefont {Knap}},\ and\ \bibinfo {author}
  {\bibfnamefont {F.}~\bibnamefont {Pollmann}},\ }\bibfield  {title} {\bibinfo
  {title} {Ergodicity breaking arising from hilbert space fragmentation in
  dipole-conserving hamiltonians},\ }\href
  {https://doi.org/10.1103/PhysRevX.10.011047} {\bibfield  {journal} {\bibinfo
  {journal} {Phys. Rev. X}\ }\textbf {\bibinfo {volume} {10}},\ \bibinfo
  {pages} {011047} (\bibinfo {year} {2020})}\BibitemShut {NoStop}%
\bibitem [{\citenamefont {Khemani}\ \emph {et~al.}(2020)\citenamefont
  {Khemani}, \citenamefont {Hermele},\ and\ \citenamefont
  {Nandkishore}}]{Khemani2020_shattering}%
  \BibitemOpen
  \bibfield  {author} {\bibinfo {author} {\bibfnamefont {V.}~\bibnamefont
  {Khemani}}, \bibinfo {author} {\bibfnamefont {M.}~\bibnamefont {Hermele}},\
  and\ \bibinfo {author} {\bibfnamefont {R.}~\bibnamefont {Nandkishore}},\
  }\bibfield  {title} {\bibinfo {title} {Localization from hilbert space
  shattering: From theory to physical realizations},\ }\href
  {https://doi.org/10.1103/PhysRevB.101.174204} {\bibfield  {journal} {\bibinfo
   {journal} {Phys. Rev. B}\ }\textbf {\bibinfo {volume} {101}},\ \bibinfo
  {pages} {174204} (\bibinfo {year} {2020})}\BibitemShut {NoStop}%
\bibitem [{\citenamefont {Rakovszky}\ \emph {et~al.}(2020)\citenamefont
  {Rakovszky}, \citenamefont {Sala}, \citenamefont {Verresen}, \citenamefont
  {Knap},\ and\ \citenamefont {Pollmann}}]{Rakovszky2020_fragmentation}%
  \BibitemOpen
  \bibfield  {author} {\bibinfo {author} {\bibfnamefont {T.}~\bibnamefont
  {Rakovszky}}, \bibinfo {author} {\bibfnamefont {P.}~\bibnamefont {Sala}},
  \bibinfo {author} {\bibfnamefont {R.}~\bibnamefont {Verresen}}, \bibinfo
  {author} {\bibfnamefont {M.}~\bibnamefont {Knap}},\ and\ \bibinfo {author}
  {\bibfnamefont {F.}~\bibnamefont {Pollmann}},\ }\bibfield  {title} {\bibinfo
  {title} {Statistical localization: From strong fragmentation to strong edge
  modes},\ }\href {https://doi.org/10.1103/PhysRevB.101.125126} {\bibfield
  {journal} {\bibinfo  {journal} {Phys. Rev. B}\ }\textbf {\bibinfo {volume}
  {101}},\ \bibinfo {pages} {125126} (\bibinfo {year} {2020})}\BibitemShut
  {NoStop}%
\bibitem [{\citenamefont {{Moudgalya}}\ \emph {et~al.}(2019)\citenamefont
  {{Moudgalya}}, \citenamefont {{Prem}}, \citenamefont {{Nandkishore}},
  \citenamefont {{Regnault}},\ and\ \citenamefont
  {{Bernevig}}}]{Moudgalya2019_arxiv}%
  \BibitemOpen
  \bibfield  {author} {\bibinfo {author} {\bibfnamefont {S.}~\bibnamefont
  {{Moudgalya}}}, \bibinfo {author} {\bibfnamefont {A.}~\bibnamefont {{Prem}}},
  \bibinfo {author} {\bibfnamefont {R.}~\bibnamefont {{Nandkishore}}}, \bibinfo
  {author} {\bibfnamefont {N.}~\bibnamefont {{Regnault}}},\ and\ \bibinfo
  {author} {\bibfnamefont {B.~A.}\ \bibnamefont {{Bernevig}}},\ }\bibfield
  {title} {\bibinfo {title} {{Thermalization and its absence within Krylov
  subspaces of a constrained Hamiltonian}},\ }\href
  {https://doi.org/10.48550/arXiv.1910.14048} {\bibfield  {journal} {\bibinfo
  {journal} {arXiv e-prints}\ ,\ \bibinfo {eid} {arXiv:1910.14048}} (\bibinfo
  {year} {2019})},\ \Eprint {https://arxiv.org/abs/1910.14048}
  {arXiv:1910.14048 [cond-mat.str-el]} \BibitemShut {NoStop}%
\bibitem [{\citenamefont {Yang}\ \emph {et~al.}(2020)\citenamefont {Yang},
  \citenamefont {Liu}, \citenamefont {Gorshkov},\ and\ \citenamefont
  {Iadecola}}]{Yang2020}%
  \BibitemOpen
  \bibfield  {author} {\bibinfo {author} {\bibfnamefont {Z.-C.}\ \bibnamefont
  {Yang}}, \bibinfo {author} {\bibfnamefont {F.}~\bibnamefont {Liu}}, \bibinfo
  {author} {\bibfnamefont {A.~V.}\ \bibnamefont {Gorshkov}},\ and\ \bibinfo
  {author} {\bibfnamefont {T.}~\bibnamefont {Iadecola}},\ }\bibfield  {title}
  {\bibinfo {title} {Hilbert-space fragmentation from strict confinement},\
  }\href {https://doi.org/10.1103/PhysRevLett.124.207602} {\bibfield  {journal}
  {\bibinfo  {journal} {Phys. Rev. Lett.}\ }\textbf {\bibinfo {volume} {124}},\
  \bibinfo {pages} {207602} (\bibinfo {year} {2020})}\BibitemShut {NoStop}%
\bibitem [{\citenamefont {De~Tomasi}\ \emph {et~al.}(2019)\citenamefont
  {De~Tomasi}, \citenamefont {Hetterich}, \citenamefont {Sala},\ and\
  \citenamefont {Pollmann}}]{tomasi2019}%
  \BibitemOpen
  \bibfield  {author} {\bibinfo {author} {\bibfnamefont {G.}~\bibnamefont
  {De~Tomasi}}, \bibinfo {author} {\bibfnamefont {D.}~\bibnamefont
  {Hetterich}}, \bibinfo {author} {\bibfnamefont {P.}~\bibnamefont {Sala}},\
  and\ \bibinfo {author} {\bibfnamefont {F.}~\bibnamefont {Pollmann}},\
  }\bibfield  {title} {\bibinfo {title} {Dynamics of strongly interacting
  systems: From fock-space fragmentation to many-body localization},\ }\href
  {https://doi.org/10.1103/PhysRevB.100.214313} {\bibfield  {journal} {\bibinfo
   {journal} {Phys. Rev. B}\ }\textbf {\bibinfo {volume} {100}},\ \bibinfo
  {pages} {214313} (\bibinfo {year} {2019})}\BibitemShut {NoStop}%
\bibitem [{\citenamefont {Roy}\ and\ \citenamefont
  {Lazarides}(2020)}]{Roy2020}%
  \BibitemOpen
  \bibfield  {author} {\bibinfo {author} {\bibfnamefont {S.}~\bibnamefont
  {Roy}}\ and\ \bibinfo {author} {\bibfnamefont {A.}~\bibnamefont
  {Lazarides}},\ }\bibfield  {title} {\bibinfo {title} {Strong ergodicity
  breaking due to local constraints in a quantum system},\ }\href
  {https://doi.org/10.1103/PhysRevResearch.2.023159} {\bibfield  {journal}
  {\bibinfo  {journal} {Phys. Rev. Res.}\ }\textbf {\bibinfo {volume} {2}},\
  \bibinfo {pages} {023159} (\bibinfo {year} {2020})}\BibitemShut {NoStop}%
\bibitem [{\citenamefont {Herviou}\ \emph {et~al.}(2021)\citenamefont
  {Herviou}, \citenamefont {Bardarson},\ and\ \citenamefont
  {Regnault}}]{Herviou2021}%
  \BibitemOpen
  \bibfield  {author} {\bibinfo {author} {\bibfnamefont {L.}~\bibnamefont
  {Herviou}}, \bibinfo {author} {\bibfnamefont {J.~H.}\ \bibnamefont
  {Bardarson}},\ and\ \bibinfo {author} {\bibfnamefont {N.}~\bibnamefont
  {Regnault}},\ }\bibfield  {title} {\bibinfo {title} {Many-body localization
  in a fragmented hilbert space},\ }\href
  {https://doi.org/10.1103/PhysRevB.103.134207} {\bibfield  {journal} {\bibinfo
   {journal} {Phys. Rev. B}\ }\textbf {\bibinfo {volume} {103}},\ \bibinfo
  {pages} {134207} (\bibinfo {year} {2021})}\BibitemShut {NoStop}%
\bibitem [{\citenamefont {Shibata}\ \emph {et~al.}(2020)\citenamefont
  {Shibata}, \citenamefont {Yoshioka},\ and\ \citenamefont
  {Katsura}}]{Shibata2020}%
  \BibitemOpen
  \bibfield  {author} {\bibinfo {author} {\bibfnamefont {N.}~\bibnamefont
  {Shibata}}, \bibinfo {author} {\bibfnamefont {N.}~\bibnamefont {Yoshioka}},\
  and\ \bibinfo {author} {\bibfnamefont {H.}~\bibnamefont {Katsura}},\
  }\bibfield  {title} {\bibinfo {title} {Onsager's scars in disordered spin
  chains},\ }\href {https://doi.org/10.1103/PhysRevLett.124.180604} {\bibfield
  {journal} {\bibinfo  {journal} {Phys. Rev. Lett.}\ }\textbf {\bibinfo
  {volume} {124}},\ \bibinfo {pages} {180604} (\bibinfo {year}
  {2020})}\BibitemShut {NoStop}%
\bibitem [{\citenamefont {Langlett}\ and\ \citenamefont
  {Xu}(2021)}]{Langlett2021}%
  \BibitemOpen
  \bibfield  {author} {\bibinfo {author} {\bibfnamefont {C.~M.}\ \bibnamefont
  {Langlett}}\ and\ \bibinfo {author} {\bibfnamefont {S.}~\bibnamefont {Xu}},\
  }\bibfield  {title} {\bibinfo {title} {Hilbert space fragmentation and exact
  scars of generalized fredkin spin chains},\ }\href
  {https://doi.org/10.1103/PhysRevB.103.L220304} {\bibfield  {journal}
  {\bibinfo  {journal} {Phys. Rev. B}\ }\textbf {\bibinfo {volume} {103}},\
  \bibinfo {pages} {L220304} (\bibinfo {year} {2021})}\BibitemShut {NoStop}%
\bibitem [{\citenamefont {Scherg}\ \emph {et~al.}(2021)\citenamefont {Scherg},
  \citenamefont {Kohlert}, \citenamefont {Sala}, \citenamefont {Pollmann},
  \citenamefont {Hebbe~Madhusudhana}, \citenamefont {Bloch},\ and\
  \citenamefont {Aidelsburger}}]{scherg2021_kinetic}%
  \BibitemOpen
  \bibfield  {author} {\bibinfo {author} {\bibfnamefont {S.}~\bibnamefont
  {Scherg}}, \bibinfo {author} {\bibfnamefont {T.}~\bibnamefont {Kohlert}},
  \bibinfo {author} {\bibfnamefont {P.}~\bibnamefont {Sala}}, \bibinfo {author}
  {\bibfnamefont {F.}~\bibnamefont {Pollmann}}, \bibinfo {author}
  {\bibfnamefont {B.}~\bibnamefont {Hebbe~Madhusudhana}}, \bibinfo {author}
  {\bibfnamefont {I.}~\bibnamefont {Bloch}},\ and\ \bibinfo {author}
  {\bibfnamefont {M.}~\bibnamefont {Aidelsburger}},\ }\bibfield  {title}
  {\bibinfo {title} {{Observing non-ergodicity due to kinetic constraints in
  tilted Fermi-Hubbard chains}},\ }\href
  {https://doi.org/10.1038/s41467-021-24726-0} {\bibfield  {journal} {\bibinfo
  {journal} {Nature Communications}\ }\textbf {\bibinfo {volume} {12}},\
  \bibinfo {pages} {4490} (\bibinfo {year} {2021})}\BibitemShut {NoStop}%
\bibitem [{\citenamefont {Feldmeier}\ and\ \citenamefont
  {Knap}(2021)}]{feldmeier2021_crit}%
  \BibitemOpen
  \bibfield  {author} {\bibinfo {author} {\bibfnamefont {J.}~\bibnamefont
  {Feldmeier}}\ and\ \bibinfo {author} {\bibfnamefont {M.}~\bibnamefont
  {Knap}},\ }\bibfield  {title} {\bibinfo {title} {Critically slow operator
  dynamics in constrained many-body systems},\ }\href
  {https://doi.org/10.1103/PhysRevLett.127.235301} {\bibfield  {journal}
  {\bibinfo  {journal} {Phys. Rev. Lett.}\ }\textbf {\bibinfo {volume} {127}},\
  \bibinfo {pages} {235301} (\bibinfo {year} {2021})}\BibitemShut {NoStop}%
\bibitem [{\citenamefont {Caldeira}\ and\ \citenamefont
  {Leggett}(1983)}]{CALDEIRA1983587}%
  \BibitemOpen
  \bibfield  {author} {\bibinfo {author} {\bibfnamefont {A.}~\bibnamefont
  {Caldeira}}\ and\ \bibinfo {author} {\bibfnamefont {A.}~\bibnamefont
  {Leggett}},\ }\bibfield  {title} {\bibinfo {title} {Path integral approach to
  quantum brownian motion},\ }\href
  {https://doi.org/https://doi.org/10.1016/0378-4371(83)90013-4} {\bibfield
  {journal} {\bibinfo  {journal} {Physica A: Statistical Mechanics and its
  Applications}\ }\textbf {\bibinfo {volume} {121}},\ \bibinfo {pages} {587}
  (\bibinfo {year} {1983})}\BibitemShut {NoStop}%
\bibitem [{\citenamefont {Buchleitner}\ \emph {et~al.}(2011)\citenamefont
  {Buchleitner}, \citenamefont {Hornberger},\ and\ \citenamefont
  {Kümmerer}}]{Buchleitner_Hornberger_Kümmerer_2011}%
  \BibitemOpen
  \bibfield  {author} {\bibinfo {author} {\bibfnamefont {A.}~\bibnamefont
  {Buchleitner}}, \bibinfo {author} {\bibfnamefont {K.}~\bibnamefont
  {Hornberger}},\ and\ \bibinfo {author} {\bibfnamefont {B.}~\bibnamefont
  {Kümmerer}},\ }\bibinfo {title} {Quantum markov processes},\ in\ \href@noop
  {} {\emph {\bibinfo {booktitle} {Coherent evolution in Noisy Environments}}}\
  (\bibinfo  {publisher} {Springer},\ \bibinfo {year} {2011})\BibitemShut
  {NoStop}%
\bibitem [{\citenamefont {Lashkari}\ \emph {et~al.}(2013)\citenamefont
  {Lashkari}, \citenamefont {Stanford}, \citenamefont {Hastings}, \citenamefont
  {Osborne},\ and\ \citenamefont {Hayden}}]{Lashkari2013}%
  \BibitemOpen
  \bibfield  {author} {\bibinfo {author} {\bibfnamefont {N.}~\bibnamefont
  {Lashkari}}, \bibinfo {author} {\bibfnamefont {D.}~\bibnamefont {Stanford}},
  \bibinfo {author} {\bibfnamefont {M.}~\bibnamefont {Hastings}}, \bibinfo
  {author} {\bibfnamefont {T.}~\bibnamefont {Osborne}},\ and\ \bibinfo {author}
  {\bibfnamefont {P.}~\bibnamefont {Hayden}},\ }\bibfield  {title} {\bibinfo
  {title} {Towards the fast scrambling conjecture},\ }\href
  {https://doi.org/10.1007/JHEP04(2013)022} {\bibfield  {journal} {\bibinfo
  {journal} {Journal of High Energy Physics}\ }\textbf {\bibinfo {volume}
  {2013}},\ \bibinfo {pages} {22} (\bibinfo {year} {2013})}\BibitemShut
  {NoStop}%
\bibitem [{\citenamefont {Zhou}\ and\ \citenamefont
  {Chen}(2019)}]{ZhouXiao2019}%
  \BibitemOpen
  \bibfield  {author} {\bibinfo {author} {\bibfnamefont {T.}~\bibnamefont
  {Zhou}}\ and\ \bibinfo {author} {\bibfnamefont {X.}~\bibnamefont {Chen}},\
  }\bibfield  {title} {\bibinfo {title} {Operator dynamics in a brownian
  quantum circuit},\ }\href {https://doi.org/10.1103/PhysRevE.99.052212}
  {\bibfield  {journal} {\bibinfo  {journal} {Phys. Rev. E}\ }\textbf {\bibinfo
  {volume} {99}},\ \bibinfo {pages} {052212} (\bibinfo {year}
  {2019})}\BibitemShut {NoStop}%
\bibitem [{\citenamefont {Xu}\ and\ \citenamefont
  {Swingle}(2019)}]{XuSwingle2019}%
  \BibitemOpen
  \bibfield  {author} {\bibinfo {author} {\bibfnamefont {S.}~\bibnamefont
  {Xu}}\ and\ \bibinfo {author} {\bibfnamefont {B.}~\bibnamefont {Swingle}},\
  }\bibfield  {title} {\bibinfo {title} {Locality, quantum fluctuations, and
  scrambling},\ }\href {https://doi.org/10.1103/PhysRevX.9.031048} {\bibfield
  {journal} {\bibinfo  {journal} {Phys. Rev. X}\ }\textbf {\bibinfo {volume}
  {9}},\ \bibinfo {pages} {031048} (\bibinfo {year} {2019})}\BibitemShut
  {NoStop}%
\bibitem [{\citenamefont {Chen}\ and\ \citenamefont
  {Zhou}(2019)}]{XiaoZhou2019}%
  \BibitemOpen
  \bibfield  {author} {\bibinfo {author} {\bibfnamefont {X.}~\bibnamefont
  {Chen}}\ and\ \bibinfo {author} {\bibfnamefont {T.}~\bibnamefont {Zhou}},\
  }\bibfield  {title} {\bibinfo {title} {Quantum chaos dynamics in long-range
  power law interaction systems},\ }\href
  {https://doi.org/10.1103/PhysRevB.100.064305} {\bibfield  {journal} {\bibinfo
   {journal} {Phys. Rev. B}\ }\textbf {\bibinfo {volume} {100}},\ \bibinfo
  {pages} {064305} (\bibinfo {year} {2019})}\BibitemShut {NoStop}%
\bibitem [{\citenamefont {Choi}(1975)}]{CHOI1975}%
  \BibitemOpen
  \bibfield  {author} {\bibinfo {author} {\bibfnamefont {M.-D.}\ \bibnamefont
  {Choi}},\ }\bibfield  {title} {\bibinfo {title} {Completely positive linear
  maps on complex matrices},\ }\href
  {https://doi.org/https://doi.org/10.1016/0024-3795(75)90075-0} {\bibfield
  {journal} {\bibinfo  {journal} {Linear Algebra and its Applications}\
  }\textbf {\bibinfo {volume} {10}},\ \bibinfo {pages} {285} (\bibinfo {year}
  {1975})}\BibitemShut {NoStop}%
\bibitem [{\citenamefont {Jamio{\l}kowski}(1972)}]{JAMIOLKOWSKI1972}%
  \BibitemOpen
  \bibfield  {author} {\bibinfo {author} {\bibfnamefont {A.}~\bibnamefont
  {Jamio{\l}kowski}},\ }\bibfield  {title} {\bibinfo {title} {Linear
  transformations which preserve trace and positive semidefiniteness of
  operators},\ }\href
  {https://doi.org/https://doi.org/10.1016/0034-4877(72)90011-0} {\bibfield
  {journal} {\bibinfo  {journal} {Reports on Mathematical Physics}\ }\textbf
  {\bibinfo {volume} {3}},\ \bibinfo {pages} {275} (\bibinfo {year}
  {1972})}\BibitemShut {NoStop}%
\bibitem [{sm()}]{sm}%
  \BibitemOpen
  \href@noop {} {}\bibinfo {note} {See Supplementary Material}\BibitemShut
  {NoStop}%
\bibitem [{Note1()}]{Note1}%
  \BibitemOpen
  \bibinfo {note} {It measures the spreading of an operator $O$ located at
  $\protect \mathbf {x}$ at time $t$ by measuring its overlap with an operator
  at $\protect \mathbf {y}$.}\BibitemShut {Stop}%
\bibitem [{mis()}]{misc2}%
  \BibitemOpen
  \href@noop {} {}\bibinfo {note} {Since $O_{\bm{x}}$ is local, the overlap
  $\lAngle \bk,\nu \Vert O \rAngle$\,$=$\,$\tr(O X_{k,\nu})$ should be near
  constant for small $k$, where $X_{k,\nu}$ is the operator corresponding to
  the Choi state $\Vert \bm{k},\nu \rAngle$.}\BibitemShut {Stop}%
\bibitem [{\citenamefont {Feynman}(1953)}]{Feynman1953}%
  \BibitemOpen
  \bibfield  {author} {\bibinfo {author} {\bibfnamefont {R.~P.}\ \bibnamefont
  {Feynman}},\ }\bibfield  {title} {\bibinfo {title} {Atomic theory of the
  $\ensuremath{\lambda}$ transition in helium},\ }\href
  {https://doi.org/10.1103/PhysRev.91.1291} {\bibfield  {journal} {\bibinfo
  {journal} {Phys. Rev.}\ }\textbf {\bibinfo {volume} {91}},\ \bibinfo {pages}
  {1291} (\bibinfo {year} {1953})}\BibitemShut {NoStop}%
\bibitem [{\citenamefont {Feynman}(1954)}]{Feynman1954}%
  \BibitemOpen
  \bibfield  {author} {\bibinfo {author} {\bibfnamefont {R.~P.}\ \bibnamefont
  {Feynman}},\ }\bibfield  {title} {\bibinfo {title} {Atomic theory of the
  two-fluid model of liquid helium},\ }\href
  {https://doi.org/10.1103/PhysRev.94.262} {\bibfield  {journal} {\bibinfo
  {journal} {Phys. Rev.}\ }\textbf {\bibinfo {volume} {94}},\ \bibinfo {pages}
  {262} (\bibinfo {year} {1954})}\BibitemShut {NoStop}%
\bibitem [{\citenamefont {Girvin}\ \emph {et~al.}(1986)\citenamefont {Girvin},
  \citenamefont {MacDonald},\ and\ \citenamefont {Platzman}}]{Girvin1986}%
  \BibitemOpen
  \bibfield  {author} {\bibinfo {author} {\bibfnamefont {S.~M.}\ \bibnamefont
  {Girvin}}, \bibinfo {author} {\bibfnamefont {A.~H.}\ \bibnamefont
  {MacDonald}},\ and\ \bibinfo {author} {\bibfnamefont {P.~M.}\ \bibnamefont
  {Platzman}},\ }\bibfield  {title} {\bibinfo {title} {Magneto-roton theory of
  collective excitations in the fractional quantum hall effect},\ }\href
  {https://doi.org/10.1103/PhysRevB.33.2481} {\bibfield  {journal} {\bibinfo
  {journal} {Phys. Rev. B}\ }\textbf {\bibinfo {volume} {33}},\ \bibinfo
  {pages} {2481} (\bibinfo {year} {1986})}\BibitemShut {NoStop}%
\bibitem [{\citenamefont {Deng}\ \emph {et~al.}(2022)\citenamefont {Deng},
  \citenamefont {Khaymovich},\ and\ \citenamefont
  {Burin}}]{deng2022superdiffusion}%
  \BibitemOpen
  \bibfield  {author} {\bibinfo {author} {\bibfnamefont {X.}~\bibnamefont
  {Deng}}, \bibinfo {author} {\bibfnamefont {I.}~\bibnamefont {Khaymovich}},\
  and\ \bibinfo {author} {\bibfnamefont {A.~L.}\ \bibnamefont {Burin}},\
  }\href@noop {} {\bibinfo {title} {Superdiffusion in random two dimensional
  system with ubiquitous long-range hopping}} (\bibinfo {year} {2022}),\
  \Eprint {https://arxiv.org/abs/2205.14715} {arXiv:2205.14715
  [cond-mat.dis-nn]} \BibitemShut {NoStop}%
\bibitem [{\citenamefont {Nandkishore}\ and\ \citenamefont
  {Hermele}(2019)}]{nandkishore2019_fractons}%
  \BibitemOpen
  \bibfield  {author} {\bibinfo {author} {\bibfnamefont {R.~M.}\ \bibnamefont
  {Nandkishore}}\ and\ \bibinfo {author} {\bibfnamefont {M.}~\bibnamefont
  {Hermele}},\ }\bibfield  {title} {\bibinfo {title} {Fractons},\ }\href
  {https://doi.org/10.1146/annurev-conmatphys-031218-013604} {\bibfield
  {journal} {\bibinfo  {journal} {Annual Review of Condensed Matter Physics}\
  }\textbf {\bibinfo {volume} {10}},\ \bibinfo {pages} {295} (\bibinfo {year}
  {2019})}\BibitemShut {NoStop}%
\bibitem [{\citenamefont {Pretko}\ \emph {et~al.}(2020)\citenamefont {Pretko},
  \citenamefont {Chen},\ and\ \citenamefont {You}}]{pretko2020fracton}%
  \BibitemOpen
  \bibfield  {author} {\bibinfo {author} {\bibfnamefont {M.}~\bibnamefont
  {Pretko}}, \bibinfo {author} {\bibfnamefont {X.}~\bibnamefont {Chen}},\ and\
  \bibinfo {author} {\bibfnamefont {Y.}~\bibnamefont {You}},\ }\bibfield
  {title} {\bibinfo {title} {Fracton phases of matter},\ }\href
  {https://doi.org/10.1142/s0217751x20300033} {\bibfield  {journal} {\bibinfo
  {journal} {International Journal of Modern Physics A}\ }\textbf {\bibinfo
  {volume} {35}},\ \bibinfo {pages} {2030003} (\bibinfo {year}
  {2020})}\BibitemShut {NoStop}%
\bibitem [{\citenamefont {Chamon}(2005)}]{chamon2005_glass}%
  \BibitemOpen
  \bibfield  {author} {\bibinfo {author} {\bibfnamefont {C.}~\bibnamefont
  {Chamon}},\ }\bibfield  {title} {\bibinfo {title} {{Quantum Glassiness in
  Strongly Correlated Clean Systems: An Example of Topological
  Overprotection}},\ }\href {https://doi.org/10.1103/PhysRevLett.94.040402}
  {\bibfield  {journal} {\bibinfo  {journal} {Phys. Rev. Lett.}\ }\textbf
  {\bibinfo {volume} {94}},\ \bibinfo {pages} {040402} (\bibinfo {year}
  {2005})}\BibitemShut {NoStop}%
\bibitem [{\citenamefont {Haah}(2011)}]{Haah11}%
  \BibitemOpen
  \bibfield  {author} {\bibinfo {author} {\bibfnamefont {J.}~\bibnamefont
  {Haah}},\ }\bibfield  {title} {\bibinfo {title} {Local stabilizer codes in
  three dimensions without string logical operators},\ }\href
  {https://doi.org/10.1103/PhysRevA.83.042330} {\bibfield  {journal} {\bibinfo
  {journal} {Phys. Rev. A}\ }\textbf {\bibinfo {volume} {83}},\ \bibinfo
  {pages} {042330} (\bibinfo {year} {2011})}\BibitemShut {NoStop}%
\bibitem [{\citenamefont {Yoshida}(2013)}]{yoshida2013_fractal}%
  \BibitemOpen
  \bibfield  {author} {\bibinfo {author} {\bibfnamefont {B.}~\bibnamefont
  {Yoshida}},\ }\bibfield  {title} {\bibinfo {title} {{Exotic topological order
  in fractal spin liquids}},\ }\href
  {https://doi.org/10.1103/PhysRevB.88.125122} {\bibfield  {journal} {\bibinfo
  {journal} {Phys. Rev. B}\ }\textbf {\bibinfo {volume} {88}},\ \bibinfo
  {pages} {125122} (\bibinfo {year} {2013})}\BibitemShut {NoStop}%
\bibitem [{\citenamefont {Vijay}\ \emph {et~al.}(2015)\citenamefont {Vijay},
  \citenamefont {Haah},\ and\ \citenamefont {Fu}}]{vijay2015_topo}%
  \BibitemOpen
  \bibfield  {author} {\bibinfo {author} {\bibfnamefont {S.}~\bibnamefont
  {Vijay}}, \bibinfo {author} {\bibfnamefont {J.}~\bibnamefont {Haah}},\ and\
  \bibinfo {author} {\bibfnamefont {L.}~\bibnamefont {Fu}},\ }\bibfield
  {title} {\bibinfo {title} {{A new kind of topological quantum order: A
  dimensional hierarchy of quasiparticles built from stationary excitations}},\
  }\href {https://doi.org/10.1103/PhysRevB.92.235136} {\bibfield  {journal}
  {\bibinfo  {journal} {Phys. Rev. B}\ }\textbf {\bibinfo {volume} {92}},\
  \bibinfo {pages} {235136} (\bibinfo {year} {2015})}\BibitemShut {NoStop}%
\bibitem [{\citenamefont {Pretko}\ and\ \citenamefont
  {Radzihovsky}(2018)}]{pretko2018_elasticity}%
  \BibitemOpen
  \bibfield  {author} {\bibinfo {author} {\bibfnamefont {M.}~\bibnamefont
  {Pretko}}\ and\ \bibinfo {author} {\bibfnamefont {L.}~\bibnamefont
  {Radzihovsky}},\ }\bibfield  {title} {\bibinfo {title} {{Fracton-Elasticity
  Duality}},\ }\href {https://doi.org/10.1103/PhysRevLett.120.195301}
  {\bibfield  {journal} {\bibinfo  {journal} {Phys. Rev. Lett.}\ }\textbf
  {\bibinfo {volume} {120}},\ \bibinfo {pages} {195301} (\bibinfo {year}
  {2018})}\BibitemShut {NoStop}%
\bibitem [{\citenamefont {Pretko}(2017)}]{PretkoSub}%
  \BibitemOpen
  \bibfield  {author} {\bibinfo {author} {\bibfnamefont {M.}~\bibnamefont
  {Pretko}},\ }\bibfield  {title} {\bibinfo {title} {{Subdimensional particle
  structure of higher rank $U(1)$ spin liquids}},\ }\href
  {https://doi.org/10.1103/PhysRevB.95.115139} {\bibfield  {journal} {\bibinfo
  {journal} {Phys. Rev. B}\ }\textbf {\bibinfo {volume} {95}},\ \bibinfo
  {pages} {115139} (\bibinfo {year} {2017})}\BibitemShut {NoStop}%
\bibitem [{\citenamefont {Pretko}(2018)}]{pretko2018_gaugprinciple}%
  \BibitemOpen
  \bibfield  {author} {\bibinfo {author} {\bibfnamefont {M.}~\bibnamefont
  {Pretko}},\ }\bibfield  {title} {\bibinfo {title} {The fracton gauge
  principle},\ }\href {https://doi.org/10.1103/PhysRevB.98.115134} {\bibfield
  {journal} {\bibinfo  {journal} {Phys. Rev. B}\ }\textbf {\bibinfo {volume}
  {98}},\ \bibinfo {pages} {115134} (\bibinfo {year} {2018})}\BibitemShut
  {NoStop}%
\bibitem [{\citenamefont {Moudgalya}\ and\ \citenamefont
  {Motrunich}(2022)}]{Moudgalya2022}%
  \BibitemOpen
  \bibfield  {author} {\bibinfo {author} {\bibfnamefont {S.}~\bibnamefont
  {Moudgalya}}\ and\ \bibinfo {author} {\bibfnamefont {O.~I.}\ \bibnamefont
  {Motrunich}},\ }\bibfield  {title} {\bibinfo {title} {Hilbert space
  fragmentation and commutant algebras},\ }\href
  {https://doi.org/10.1103/PhysRevX.12.011050} {\bibfield  {journal} {\bibinfo
  {journal} {Phys. Rev. X}\ }\textbf {\bibinfo {volume} {12}},\ \bibinfo
  {pages} {011050} (\bibinfo {year} {2022})}\BibitemShut {NoStop}%
\bibitem [{\citenamefont {Iaconis}\ \emph {et~al.}(2021)\citenamefont
  {Iaconis}, \citenamefont {Lucas},\ and\ \citenamefont
  {Nandkishore}}]{iaconis2021_multipole}%
  \BibitemOpen
  \bibfield  {author} {\bibinfo {author} {\bibfnamefont {J.}~\bibnamefont
  {Iaconis}}, \bibinfo {author} {\bibfnamefont {A.}~\bibnamefont {Lucas}},\
  and\ \bibinfo {author} {\bibfnamefont {R.}~\bibnamefont {Nandkishore}},\
  }\bibfield  {title} {\bibinfo {title} {{Multipole conservation laws and
  subdiffusion in any dimension}},\ }\href
  {https://doi.org/10.1103/PhysRevE.103.022142} {\bibfield  {journal} {\bibinfo
   {journal} {Phys. Rev. E}\ }\textbf {\bibinfo {volume} {103}},\ \bibinfo
  {pages} {022142} (\bibinfo {year} {2021})}\BibitemShut {NoStop}%
\bibitem [{\citenamefont {Singh}\ \emph {et~al.}(2021)\citenamefont {Singh},
  \citenamefont {Ware}, \citenamefont {Vasseur},\ and\ \citenamefont
  {Friedman}}]{Singh_2021}%
  \BibitemOpen
  \bibfield  {author} {\bibinfo {author} {\bibfnamefont {H.}~\bibnamefont
  {Singh}}, \bibinfo {author} {\bibfnamefont {B.~A.}\ \bibnamefont {Ware}},
  \bibinfo {author} {\bibfnamefont {R.}~\bibnamefont {Vasseur}},\ and\ \bibinfo
  {author} {\bibfnamefont {A.~J.}\ \bibnamefont {Friedman}},\ }\bibfield
  {title} {\bibinfo {title} {Subdiffusion and many-body quantum chaos with
  kinetic constraints},\ }\bibfield  {journal} {\bibinfo  {journal} {Physical
  Review Letters}\ }\textbf {\bibinfo {volume} {127}},\ \href
  {https://doi.org/10.1103/physrevlett.127.230602}
  {10.1103/physrevlett.127.230602} (\bibinfo {year} {2021})\BibitemShut
  {NoStop}%
\bibitem [{\citenamefont {Feng}\ and\ \citenamefont
  {Skinner}(2022)}]{skinner2022fracton}%
  \BibitemOpen
  \bibfield  {author} {\bibinfo {author} {\bibfnamefont {X.}~\bibnamefont
  {Feng}}\ and\ \bibinfo {author} {\bibfnamefont {B.}~\bibnamefont {Skinner}},\
  }\bibfield  {title} {\bibinfo {title} {Hilbert space fragmentation produces
  an effective attraction between fractons},\ }\href
  {https://doi.org/10.1103/PhysRevResearch.4.013053} {\bibfield  {journal}
  {\bibinfo  {journal} {Phys. Rev. Res.}\ }\textbf {\bibinfo {volume} {4}},\
  \bibinfo {pages} {013053} (\bibinfo {year} {2022})}\BibitemShut {NoStop}%
\bibitem [{\citenamefont {Zechmann}\ \emph {et~al.}(2022)\citenamefont
  {Zechmann}, \citenamefont {Altman}, \citenamefont {Knap},\ and\ \citenamefont
  {Feldmeier}}]{zechmann2022fractonic}%
  \BibitemOpen
  \bibfield  {author} {\bibinfo {author} {\bibfnamefont {P.}~\bibnamefont
  {Zechmann}}, \bibinfo {author} {\bibfnamefont {E.}~\bibnamefont {Altman}},
  \bibinfo {author} {\bibfnamefont {M.}~\bibnamefont {Knap}},\ and\ \bibinfo
  {author} {\bibfnamefont {J.}~\bibnamefont {Feldmeier}},\ }\bibfield  {title}
  {\bibinfo {title} {Fractonic luttinger liquids and supersolids in a
  constrained bose-hubbard model},\ }\href@noop {} {\bibfield  {journal}
  {\bibinfo  {journal} {arXiv preprint arXiv:2210.11072}\ } (\bibinfo {year}
  {2022})}\BibitemShut {NoStop}%
\bibitem [{\citenamefont {Horn}(1981)}]{horn1981finite}%
  \BibitemOpen
  \bibfield  {author} {\bibinfo {author} {\bibfnamefont {D.}~\bibnamefont
  {Horn}},\ }\bibfield  {title} {\bibinfo {title} {Finite matrix models with
  continuous local gauge invariance},\ }\href@noop {} {\bibfield  {journal}
  {\bibinfo  {journal} {Physics Letters B}\ }\textbf {\bibinfo {volume}
  {100}},\ \bibinfo {pages} {149} (\bibinfo {year} {1981})}\BibitemShut
  {NoStop}%
\bibitem [{\citenamefont {Orland}\ and\ \citenamefont
  {Rohrlich}(1990)}]{orland1990lattice}%
  \BibitemOpen
  \bibfield  {author} {\bibinfo {author} {\bibfnamefont {P.}~\bibnamefont
  {Orland}}\ and\ \bibinfo {author} {\bibfnamefont {D.}~\bibnamefont
  {Rohrlich}},\ }\bibfield  {title} {\bibinfo {title} {Lattice gauge magnets:
  Local isospin from spin},\ }\href@noop {} {\bibfield  {journal} {\bibinfo
  {journal} {Nuclear Physics B}\ }\textbf {\bibinfo {volume} {338}},\ \bibinfo
  {pages} {647} (\bibinfo {year} {1990})}\BibitemShut {NoStop}%
\bibitem [{\citenamefont {Chandrasekharan}\ and\ \citenamefont
  {Wiese}(1997)}]{chandrasekharan1997quantum}%
  \BibitemOpen
  \bibfield  {author} {\bibinfo {author} {\bibfnamefont {S.}~\bibnamefont
  {Chandrasekharan}}\ and\ \bibinfo {author} {\bibfnamefont {U.-J.}\
  \bibnamefont {Wiese}},\ }\bibfield  {title} {\bibinfo {title} {Quantum link
  models: A discrete approach to gauge theories},\ }\href@noop {} {\bibfield
  {journal} {\bibinfo  {journal} {Nuclear Physics B}\ }\textbf {\bibinfo
  {volume} {492}},\ \bibinfo {pages} {455} (\bibinfo {year}
  {1997})}\BibitemShut {NoStop}%
\bibitem [{\citenamefont {Fradkin}\ and\ \citenamefont
  {Kivelson}(1990)}]{fradkin1990short}%
  \BibitemOpen
  \bibfield  {author} {\bibinfo {author} {\bibfnamefont {E.}~\bibnamefont
  {Fradkin}}\ and\ \bibinfo {author} {\bibfnamefont {S.}~\bibnamefont
  {Kivelson}},\ }\bibfield  {title} {\bibinfo {title} {Short range resonating
  valence bond theories and superconductivity},\ }\href@noop {} {\bibfield
  {journal} {\bibinfo  {journal} {Modern Physics Letters B}\ }\textbf {\bibinfo
  {volume} {4}},\ \bibinfo {pages} {225} (\bibinfo {year} {1990})}\BibitemShut
  {NoStop}%
\bibitem [{\citenamefont {Moessner}\ \emph {et~al.}(2001)\citenamefont
  {Moessner}, \citenamefont {Sondhi},\ and\ \citenamefont
  {Fradkin}}]{moessner_rvb}%
  \BibitemOpen
  \bibfield  {author} {\bibinfo {author} {\bibfnamefont {R.}~\bibnamefont
  {Moessner}}, \bibinfo {author} {\bibfnamefont {S.~L.}\ \bibnamefont
  {Sondhi}},\ and\ \bibinfo {author} {\bibfnamefont {E.}~\bibnamefont
  {Fradkin}},\ }\bibfield  {title} {\bibinfo {title} {Short-ranged resonating
  valence bond physics, quantum dimer models, and ising gauge theories},\
  }\href {https://doi.org/10.1103/PhysRevB.65.024504} {\bibfield  {journal}
  {\bibinfo  {journal} {Phys. Rev. B}\ }\textbf {\bibinfo {volume} {65}},\
  \bibinfo {pages} {024504} (\bibinfo {year} {2001})}\BibitemShut {NoStop}%
\bibitem [{\citenamefont {Lake}\ \emph {et~al.}(2022)\citenamefont {Lake},
  \citenamefont {Lee}, \citenamefont {Han},\ and\ \citenamefont
  {Senthil}}]{lake2022dipole}%
  \BibitemOpen
  \bibfield  {author} {\bibinfo {author} {\bibfnamefont {E.}~\bibnamefont
  {Lake}}, \bibinfo {author} {\bibfnamefont {H.-Y.}\ \bibnamefont {Lee}},
  \bibinfo {author} {\bibfnamefont {J.~H.}\ \bibnamefont {Han}},\ and\ \bibinfo
  {author} {\bibfnamefont {T.}~\bibnamefont {Senthil}},\ }\bibfield  {title}
  {\bibinfo {title} {Dipole condensates in tilted bose-hubbard chains},\
  }\href@noop {} {\bibfield  {journal} {\bibinfo  {journal} {arXiv preprint
  arXiv:2210.02470}\ } (\bibinfo {year} {2022})}\BibitemShut {NoStop}%
\bibitem [{\citenamefont {Morningstar}\ \emph {et~al.}(2023)\citenamefont
  {Morningstar}, \citenamefont {O'Dea},\ and\ \citenamefont
  {Richter}}]{PhysRevB.108.L020304}%
  \BibitemOpen
  \bibfield  {author} {\bibinfo {author} {\bibfnamefont {A.}~\bibnamefont
  {Morningstar}}, \bibinfo {author} {\bibfnamefont {N.}~\bibnamefont {O'Dea}},\
  and\ \bibinfo {author} {\bibfnamefont {J.}~\bibnamefont {Richter}},\
  }\bibfield  {title} {\bibinfo {title} {Hydrodynamics in long-range
  interacting systems with center-of-mass conservation},\ }\href
  {https://doi.org/10.1103/PhysRevB.108.L020304} {\bibfield  {journal}
  {\bibinfo  {journal} {Phys. Rev. B}\ }\textbf {\bibinfo {volume} {108}},\
  \bibinfo {pages} {L020304} (\bibinfo {year} {2023})}\BibitemShut {NoStop}%
\bibitem [{\citenamefont {Gliozzi}\ \emph {et~al.}(2023)\citenamefont
  {Gliozzi}, \citenamefont {May-Mann}, \citenamefont {Hughes},\ and\
  \citenamefont {De~Tomasi}}]{PhysRevB.108.195106}%
  \BibitemOpen
  \bibfield  {author} {\bibinfo {author} {\bibfnamefont {J.}~\bibnamefont
  {Gliozzi}}, \bibinfo {author} {\bibfnamefont {J.}~\bibnamefont {May-Mann}},
  \bibinfo {author} {\bibfnamefont {T.~L.}\ \bibnamefont {Hughes}},\ and\
  \bibinfo {author} {\bibfnamefont {G.}~\bibnamefont {De~Tomasi}},\ }\bibfield
  {title} {\bibinfo {title} {Hierarchical hydrodynamics in long-range
  multipole-conserving systems},\ }\href
  {https://doi.org/10.1103/PhysRevB.108.195106} {\bibfield  {journal} {\bibinfo
   {journal} {Phys. Rev. B}\ }\textbf {\bibinfo {volume} {108}},\ \bibinfo
  {pages} {195106} (\bibinfo {year} {2023})}\BibitemShut {NoStop}%
\bibitem [{\citenamefont {Ogunnaike}\ \emph {et~al.}(2023)\citenamefont
  {Ogunnaike}, \citenamefont {Feldmeier},\ and\ \citenamefont {Lee}}]{prev}%
  \BibitemOpen
  \bibfield  {author} {\bibinfo {author} {\bibfnamefont {O.}~\bibnamefont
  {Ogunnaike}}, \bibinfo {author} {\bibfnamefont {J.}~\bibnamefont
  {Feldmeier}},\ and\ \bibinfo {author} {\bibfnamefont {J.~Y.}\ \bibnamefont
  {Lee}},\ }\href@noop {} {\bibinfo {title} {Unifying emergent hydrodynamics
  and lindbladian low energy spectra across symmetries, constraints, and
  long-range interactions}} (\bibinfo {year} {2023}),\ \Eprint
  {https://arxiv.org/abs/2304.13028v1} {arXiv:2304.13028v1 [cond-mat.str-el]}
  \BibitemShut {NoStop}%
\bibitem [{Note2()}]{Note2}%
  \BibitemOpen
  \bibinfo {note} {This means we have translation symmetry for left and right
  Hilbert space in the doubled Hilbert space formalism}\BibitemShut {NoStop}%
\end{thebibliography}%

\newpage 

\appendix

\begin{center}
    \bf{SUPPLEMENTARY MATERIALS}
\end{center}

\tableofcontents

\section{Doubled Hilbert Space Formalism} \label{app:Choi}
 
In this section, we provide a brief introduction to the Choi isomorphism and doubled Hilbert space formalism. For a given Hermitian operator $O =\sum_j \lambda_j |\psi_j \rangle \langle \psi_j|$ acting on the Hilbert space ${\cal H}$ (such as a physical observable or density matrix), the Choi state $\Vert O \rAngle$~\cite{CHOI1975, JAMIOLKOWSKI1972} is defined in a doubled Hilbert space ${\cal H}_d={\cal H}_u \otimes {\cal H}_l$ as follows:
\begin{align}
        \Vert \mathbb{I} \rAngle &:= \sum_i |i \rangle_u \otimes   |i\rangle_l \nonumber \\
        \Vert O \rAngle &:= (\mathbb{I} \otimes O) \Vert \mathbb{I} \rAngle  = \sum_i |i \rangle_u \otimes O |i\rangle_l \nonumber \\
        &=  \sum_f \lambda_j  | \psi^*_j \rangle_u \otimes | \psi_j \rangle_l = (O^T \otimes \mathbb{I}) \Vert \mathbb{I} \rAngle,
\end{align}
where subscripts $u,l$ is introduced to distinguish two copies of $\cH$. Also, note that $(A \otimes B) \Vert \mathbb{I} \rAngle = (A B^T \otimes \mathbb{I}) \Vert \mathbb{I} \rAngle = (\mathbb{I} \otimes BA^T) \Vert \mathbb{I} \rAngle $. Therefore, under Choi Isomorphism, $ A O B \mapsto (B^T \otimes A) \Vert O \rAngle$.
Note that, for a given Choi state $\Vert O \rAngle$, its operator form can then be obtained by taking an overlap with a states $\Vert i,j \rAngle \equiv \vert i \rangle \otimes \vert j \rangle \in \cH_d$:
\begin{align}
    \lAngle i,j \Vert O \rAngle=\langle j | O | i \rangle.
\end{align}

The Choi state automatically respects the following symmetry:
\begin{align}
    \textrm{SWAP}^* \equiv {\cal C} \circ \textrm{SWAP},
\end{align}
where the SWAP symmetry exchanges ${\cal H}_u$ and ${\cal H}_l$, and ${\cal C}$ is the complex-conjugation symmetry. This operation corresponds to Hermitian conjugation in the operator language. 

Similarly, under the Choi isomorphism, a quantum channel acting on the space of linear operators defined on $\cH$ would map into a linear operator (not necessarily Hermitian) defined on $\cH_d$, namely the \emph{Choi operator}. For a generic quantum channel $\cE$ with Kraus representation $\{ K_i \}$ ($\cE: \rho \mapsto \sum_i K_i \rho K_i^\dagger$), its Choi operator form is defined as the following:
\begin{align}
    \cE \mapsto \hat{\cE} \equiv \sum_i K_i^* \otimes K_i.
\end{align}
For example, the averaged action of the Brownian time evolution in Eq.(\textcolor{red}{1}) from the main text would be mapped into a Lindbladian operator acting on the doubled Hilbert space under the Choi isomorphism.

Turning to symmetry constraints, let $G$ be the symmetry group acting on the original Hilbert space $\cH$. Then, the Choi state will enjoy a doubled symmetry group $G_u \times G_l$. We remark that the symmetry representation of $g \in G$ in the upper Hilbert space, ${\cal H}_u$, is defined as a complex-conjugated version of the original representation, $U^*(g)$. Accordingly, 
\begin{align}
    |\Psi \rangle \mapsto (U^*(g_u) \otimes U(g_l)) |\Psi \rangle \quad \forall g_u \cdot g_l \in G_u \times G_l.
\end{align}

\section{Brownian Circuit and the Lindbladian} \label{app:Lind}

In the body of this paper, we chose a specific type of Brownian circuit in order to demonstrate charge transport in a clear manner. Here, we will derive the Choi operator for the averaged dynamics of a general Brownian circuit. The most general Brownian circuit employs random variables $\{dB_i\}$ for the timeslice $[t,t+\Delta)$ such that the first moment $\mathbb{E}[dB_i]=\mu_i$ and the second moment $\mathbb{E}[dB_i dB_j ]=\mu_i \mu_j + \delta_{ij}/ \Delta $. Using these variables, the Hamiltonian at time slice $[t,t+\Delta)$ is defined as $H_t \equiv \sum_i h_i dB_{i,t}$, so that a density matrix $\rho_t$ evolves as
\begin{align}
    &e^{-i H_t \Delta} \rho_t  e^{i H_t \Delta}=\rho_t - i \Delta  \sum_i [h_i, \rho_t] dB_i \nonumber \\
    & \quad \qquad \quad - \frac{\Delta^2}{2} \sum_{i,j} [h_i, [h_j, \rho_t]] dB_i\,dB_j + \cdots  .
\end{align}
This allows one to characterize the expected continuous-time dynamics of $\rho_t$ as
\begin{align}
   & \mathbb{E}[ \rd_\tau \rho ] \equiv \lim_{\Delta \rightarrow 0} \frac{  \mathbb{E}[\rho_{t+\Delta} - \rho_t]}{\Delta} \nonumber \\
   & \quad=\sum_i \Big( - i\mu_i[h_i,\rho]- \frac{1}{2} [h_i, [h_i, \rho]] \Big) \nonumber  \\
   & \quad=\sum_i \Big( - i\mu_i(h_i\rho - \rho h_i) - \frac{1}{2} ( h_i^2 \rho - 2 h_i \rho h_i + \rho h_i^2)  \Big).
\end{align}
Employing the Choi isomorphism explained in the previous section, we can recast the operator $\rho_t$ as a state vector in a doubled Hilbert space, $\Vert \rho_t \rAngle$. Similarly, the above action of averaged time evolution, which can be understood as a quantum channel, can be recast into a linear operator $\hat{ {\cal H} }_{\cal L}$ acting on the doubled state as $\rd_\tau \Vert \rho \rAngle = - \hat{\cH}_\cL \Vert \rho \rAngle$  where
\begin{align}
\label{Eq:General_Brownian}
    \hat{ {\cal H} }_{\cal L} &= \sum_i i\mu_i\qty(h_i^T \otimes \mathbb{I} -  \mathbb{I} \otimes h_i) + \qty(h_i^T \otimes \mathbb{I} -  \mathbb{I} \otimes h_i )^2 \nonumber \\ 
    &=\sum_{i} \big( i\mu_{i}{\cal O}^\vdagger_{i} + {\cal O}^\dagger_{i}  {\cal O}^\vdagger_{i} \big),
\end{align}
where $\cO_{i} \equiv h_i^T \otimes \mathbb{I} - \mathbb{I} \otimes h_i$. At $\mu_{i}=0$, we recover Eq.(\textcolor{red}{3}) in the main text. 

In fact, a similar structure is obtained from the master equation in Lindblad from, where the time evolution of the density matrix (or operator) is given as
\begin{align}\label{Eq.fullEffHam}
   \dot{\rho}= -i[H,\rho]+  \sum_i\gamma_i\qty( L_i\rho L_i^\dagger -  \frac{1}{2}\{ L_i^\dagger L_i, \rho\}).
\end{align}
Under the Choi isomorphism, the RHS can be expressed as the action of the following linear operator $\hat{\cH}_\cL$ on $\Vert \rho \rAngle$:
\begin{align}
    \hat{ {\cal H} }_{\cal L} &= - i\qty(H^T \otimes \mathbb{I} - \mathbb{I} \otimes H ) \nonumber \\
     & - \sum_i \frac{\gamma_i}{2} \qty(2L_i^* \otimes L_i -  (L_i^\dagger L_i)^T \otimes \mathbb{I} -\mathbb{I} \otimes L_i^\dagger L_i ),
\end{align}
where $H$ is the system Hamiltonian, $L_i$ are the jump operators, and $\gamma_i \geq 0$ are the damping weights. When the jump operators are Hermitian up to a phase, i.e. $L_i^\dagger=e^{i\theta} L_i$, and we set $H=0$, we see a familiar form:
 \begin{align}
     \hat{ {\cal H} }_{\cal L} &=  \sum_i \frac{\gamma_i}{2} | L_i^T \otimes \mathbb{I}  -\mathbb{I} \otimes L_i |^2 = \frac{1}{2}\sum_{{\bm{x}},\nu}  {\cal \tilde{O}}^\dagger_{{\bm{x}},\nu}  {\cal \tilde{O}}^\vdagger_{{\bm{x}},\nu},
\end{align}
where ${\tilde{O}}^\vdagger_{i = ({\bm{x}},\nu)}= L_i^T \otimes \mathbb{I}  -\mathbb{I} \otimes L_i$. Thus, in a system obeying Lindbladian dynamics governed by hermitian jump operators, our results should hold. The intuition about why such a system would imitate random Brownian evolution comes from the fact that these conditions imply that the relevant system dynamics all come from interactions with an infinite-temperature environmental bath.

\section{Generalization}

In the main body of the paper, we have discussed the Brownian dynamics essentially captured by \eqnref{Eq.fullEffHam} with $H=0$ and $L = e^{i\theta}L^\dagger$. 
In the following, we relax this condition and investigate how this gives rise to non-Hermitian effective Hamiltonian. 

\subsection{Discrete Evolution}

First, we remark that the averaged dynamics in the doubled Hilbert space can be expressed as
\begin{align}
    e^{-\hat{\cH}_{\cL} dt} := \int d B_t p(dB_t) \big( e^{-i h^T(dB_t) dt} \otimes e^{-i h(dB_t) dt} \big) 
\end{align}
where $p(dB_t)$ is the probability distribution for the Brownian random variable $dB_t$. The above expression can be immediately extended to any random local-unitary ensemble ${\cal U} = \{ (p(U), U) \}$ where each $U$ conserves a desired symmetry (e.g. $\U(1)$) by replacing a continuous time evolution by a discrete-time evolution, where its averaged dynamics for each time step is captured as
\begin{align}
    e^{-\cH } = \int dU p(U) \big( U^T \otimes U \big).
\end{align}
Note that the RHS has its eigenvalue magnitudes always equal or smaller than $1$, implying that $\cH \succcurlyeq 0$. Now, the late-time averaged dynamics of this random ensemble ${\cal U}$ is captured by the low energy spectrum of the effective Hamiltonian $\cH$, since after the application of this random unitary circuit layers $T \gg 1$ times, the Choi states with small eigenvalues against $\cH$ would survive.

\subsection{Quantum Coherent Terms}

Second, for more generic Lindbladian evolution, we loosen the restrictions imposed by Brownian evolution and add quantum-coherent terms to our Lindbladian to examine the effects on our dynamics.  To do so, we return to \eqnref{Eq:General_Brownian} with $\mu_i \neq 0$ or \eqnref{Eq.fullEffHam} with $H \neq 0$. Treating each term separately, we get
\begin{align}\label{quantumdecompose}
    \hat{ {\cal H} }_{\cal L} &= i H_1 + H_2 
  \nonumber \\
    H_1 &=  -\qty(H^T \otimes \mathbb{I} - \mathbb{I} \otimes H )  \nonumber \\ 
    H_2 &= \sum_i\frac{\gamma_i}{2} \left( L_i^T \otimes \mathbb{I} - \mathbb{I} \otimes 
    L_i \right)^\dagger \left( L_i^T \otimes \mathbb{I} - \mathbb{I}\otimes L_i \right)\nonumber \\
\end{align}
where $H_1$ and $H_2$ are Hermitian operators. Note that $H_2$ is positive semi-definite and  $\Vert \mathbb{I} \rAngle$ is the ground state with zero energy, i.e., $H_{1} \Vert \mathbb{I} \rAngle = H_{2} \Vert \mathbb{I} \rAngle  = 0$. This is because
\begin{align}\label{switch}
    (O \otimes \mathbb{I} ) \Vert \mathbb{I} \rAngle = (\mathbb{I} \otimes O^T) \Vert \mathbb{I} \rAngle.
\end{align}
When $H_2 = 0$, and $H_1$ respects translational symmetry \emph{strongly}~\footnote{This means we have translation symmetry for left and right Hilbert space in the doubled Hilbert space formalism}, the evolution is entirely coherent, and we would expect charge dynamics to be ballistic due to momentum conservation.
However, when $H_1 = 0$, as shown in the main text, charge dynamics would be diffusive.

Although we cannot apply variational estimates for the non-Hermitian Hamiltonian, let us attempt to understand what would happen if both coherent $(H_1)$ and stochastic $(H_2$) terms are present. We would assume that each component is independently symmetry preserving: $[H_{1,2},Q_{\textrm{diag}}] = 0$ (this condition is equivalent to $[H,Q] = [L_i,Q] = 0$).  These two components may not commute; however, due to translation invariance, they should still have a spectrum parameterized by momentum eigenstates as $E_{\bm{k},\nu} = if_1({\bm{k}},\nu) + f_2(\bm{k},\nu)$. As such, the autocorrelation function should take the form:
\begin{align}
    &\mathbb{E}  \langle O_{{\bm{y}}}(t) O_{{\bm{x}}}(0) \rangle_\rho \nonumber \\
    &\quad \propto \sum_{{\bm{k}},\nu}  e^{i{\bm{k}} \cdot \Delta {\bm{x}}- if_1(\bm{k},\nu)t} e^{-t f_2(\bm{k},\nu)}| \lAngle {\bm{k}},\nu \Vert O_{{\bm{x}}} \rAngle|^2
\end{align}
where $\Delta x = y - x$. Now if we assume that $f_1(\bm{k},\nu) \approx c_\nu\bm{k} + \dots$ is approximately linear for small k, and $f_2(\bm{k},\nu) \sim k^{\beta}$, then we obtain just what we might expect: a decay determined by the real part, yielding  an expected decay, but with one operator in the autocorrelation shifted by distance $\Delta x  = c_\nu t$
\begin{align} 
    \mathbb{E}  \langle O_{{\bm{x-c_\nu t}}}(t) O_{{\bm{x}}}(0)\rangle_\rho  \underset{t \rightarrow \infty}{\sim}  \int_k e^{-t (E_0+k^\beta)} \dd^d \bk \sim \frac{e^{-t E_0}}{|t|^{d/\beta}}.
\end{align}
Now, let us evaluate real and imaginary components using variational states defined in the main text: 
\begin{align} 
    \lAngle m_{\bm{k}} \Vert \hat{ {\cal H} }_{\cal L} \Vert m_{\bm{k}} \rAngle &  = i \lAngle m_{\bm{k}} \Vert H_1\Vert m_{\bm{k}} \rAngle + \lAngle m_{\bm{k}} \Vert H_2\Vert m_{\bm{k}} \rAngle \nonumber \\
    & = i \lAngle m \Vert \rho_{-k}[H_1,\rho_k]\Vert m \rAngle  \nonumber \\
    &+ \sum_{i,\lambda} \lAngle m\Vert [\cO_{i,\lambda},\rho_k]^\dagger[\cO_{i,\lambda},\rho_k] \Vert m \rAngle \nonumber \\
    &\underset{k\rightarrow 0}{\approx} iC_1k + C_2 k^2
\end{align}
The first term is linear in $k$ for a generic $H_1$  because $ [H_1,\rho_k] = \sum_{n,x} \frac{(ikx)^n}{n!}[H_1,\rho_x]$, and the $n=0$ term vanishes due to symmetry.  
Naively, this would seem to give a ballistically spreading front; however, 
we have to be careful when applying our variational estimate.  While it is still true that our variational modes may bound the real spectrum arising from $H_2$, the imaginary component coming from $H_1$ should not have any such bound. Instead, the imaginary component of the spectrum will generically either be gapped or have a linear dispersion. Expanding to the lowest order, we have $f_1(\bm{k},\nu) \approx c_{0,\nu} + c_\nu\bm{k} + \dots$, where $c_{0,\nu}$ can be zero.

A non-zero $c_{0,\nu}$ is generically expected, resulting in diffusion with some form of oscillating phase factor:
\begin{align} 
    \mathbb{E}  \langle O_{{\bm{y}}}(t) O_{{\bm{x}}}(0)\rangle_\rho  & \underset{t \rightarrow \infty}{\sim}  \int_k e^{i(k\Delta x-C_1t)}e^{-t C_2k^2} d k \\ \nonumber
    & \sim e^{- iC_1t}\frac{e^{-\frac{(\Delta x )^2}{4C_2t} }}{\sqrt{C_2 t}}.
\end{align}
To make this more precise, we need to bring up an important fact about Lindbladians.  Because they preserve hermiticity (as encoded by the SWAP symmetry in Sec.\,A, which is complex-conjugation symmetry in the operator formalism), any complex-valued eigenstates of our effective Hamiltonian, $\hat{ {\cal H} }_{\cal L}$, must come in conjugate pairs. As such, there will always exist an index $\nu'$ such that, at lowest order, the spectrum is $E_{\bm{k},\nu'} = iC_{1,\nu'} + C_{2,\nu'}k^2$, where $C_{1,\nu'} = -C_{1,\nu}$, $C_{2,\nu'} = C_{2,\nu}$.  Thus, we obtain multiple spreading fronts, so that, summing over these pairs 
\begin{align}
    \mathbb{E}  \langle O_{{\bm{y}}}(t) O_{{\bm{x}}}(0)\rangle_\rho  \underset{t \rightarrow \infty}{\sim}  \sum_{\nu \neq \nu'} \frac{\cos{(C_{1,\nu} t)}e^{-\frac{(\Delta x)^2}{4C_{2,\nu}t} } }{\sqrt{C_{2,\nu} t}},
\end{align}
which gives a diffusive dynamics with an oscillating prefactor.

\subsection{Non-Hermitian Jump Operators}

Finally, what if $L_i^\dagger \neq e^{i \theta} L_i$ for any $\theta$? To proceed, we have to understand the non-Hermitian nature of $\hat{\cH}_\cL$ better. As a non-Hermitian Hamiltonian, $\hat{\mathcal{H}_{\lind}}$ will generically have different left and right eigenvectors $\lAngle L_\alpha \Vert$ 
 and $\Vert R_\alpha \rAngle$ for the shared eigenvalue $\epsilon_\alpha$. Although generically right (left) eigenvectors are not orthogonal to each other, they satisfy a biorthonormality:
 \begin{align}
     \lAngle L_\alpha \Vert R_\beta \rAngle = \delta_{\alpha \beta}.
 \end{align}
Thus, if we wish to describe to the evolution of an operator $\langle O(t) \rangle = \textrm{tr}(O \rho(t)) =  \lAngle O \Vert e^{-t \hat{\mathcal{H}_{\lind}}} \Vert \rho \rAngle$, we may choose to study late-time dynamics in the Schrodinger picture by focusing on right eigenvectors with ground state $\Vert \rho_{eq}\rAngle$, or in the Heisenberg picture by focusing on left eigenvectors with ground state $\lAngle \mathbb{I} \Vert$.

To carry out further analysis, we switch from density matrix evolution to operator evolution. This is because the steady state for density matrix evolution is generically not the maximally mixed state $\mathbb{I}$, yet that of operator evolution always is due to the trace-preserving nature of the Lindbladian dynamics. 
\begin{align}
    &\textrm{tr}( \rho(t) ) = \lAngle \mathbb{I} \Vert \rho(t) \rAngle =  1 \nonumber \\
    \Rightarrow & \quad \partial_t \lAngle \mathbb{I} \Vert \rho \rAngle = \lAngle \mathbb{I} \Vert \hat{\mathcal{H}_{\lind}} \Vert \rho \rAngle = 0
\end{align}
Since this holds for any density matrix, $\Vert \rho \rAngle$, it is clear that $\lAngle \mathbb{I} \Vert$ is a left-ground state. This ground state structure greatly simplifies our analysis, and since we care about the rates of decay, not the exact form of the equilibrium density matrix, we focus on operator evolution.

The operator evolution is given as
\begin{align}
   \dot{O} &= i[H,O]+  \sum_i \gamma_i \qty( L_i^\dagger O L_i -  \frac{1}{2}\{ L_i^\dagger L_i, O\}) \nonumber \\
   & = i[H,O]+ \sum_i \frac{\gamma_i}{2} \left( \{L^{h}_i,[O,L^{a}_i] \} - \{ L^{a}_i,[O,L^{h}_i]\}\right) \nonumber \\
   & + \sum_i \frac{\gamma_i}{2}\left( [L^{h}_i,[O,L^{h}_i]] + [L^{a}_i,[O,L^{a}_i]]\right)
\end{align}
Where we have split the jump operator into hermitian and antihermitian components, $L_i = L_i^{h} + iL_i^{a}$, where both $L^{h}$ and $L^{a}$ are hermitian. Using the Choi-Isomorphism, this can be translated into an effective Hamiltonian of the form
\begin{align}\label{fulldecompose}
    \hat{\mathcal{H}_\lind} &=  i H_1 + H_2 \nonumber \\
    H_1 &= \qty(H^T \otimes \mathbb{I} - \mathbb{I} \otimes H ) \nonumber \\
    & + \sum_i\frac{\gamma_i}{2} \qty( (L_i^{h})^T \otimes \mathbb{I} + \mathbb{I} \otimes 
    L_i^{h} ) \qty(\mathbb{I}\otimes (L_i^{a})^T -  L_i^{a} \otimes \mathbb{I} ) \nonumber \\
    & - \sum_i\frac{\gamma_i}{2} \qty( (L_i^{a})^T \otimes \mathbb{I} + \mathbb{I} \otimes 
    L_i^{a} ) \qty(\mathbb{I}\otimes (L_i^{h})^T -  L_i^{h} \otimes \mathbb{I} ) \nonumber \\
     H_2 &= \sum_i \frac{\gamma_i}{2} \qty((L_i^{h})^T \otimes \mathbb{I}  -\mathbb{I} \otimes L_i^{h} )^2 \nonumber\\
     &+\sum_i \frac{\gamma_i}{2} \qty((L_i^{a})^T \otimes \mathbb{I}  -\mathbb{I} \otimes L_i^{a} )^2,
\end{align}
where $H_1$ and $H_2$ are both Hermitian.
Using \eqref{switch}, it quickly follows that $H_{1} \Vert \mathbb{I} \rAngle = H_{2} \Vert \mathbb{I} \rAngle  = 0$. Given this, using the same logic as above, we see that the new $iH_1$ will generically contribute to an oscillating amplitude, now with additional influence from the jump operators $L_i$.  Naively applying the variational estimate from the previous subsection, $H_2$, the sum of two locally squared operators, will once more result in diffusion. However, a rigorous understanding of diffusive dynamics with this non-Hermitian effective Hamiltonian would require further study.

 \section{Feynman-Bijl Formula} \label{app:FB}

The collective excitations described in the body of this paper closely mirror variational density fluctuation modes in bosonic systems, as described by the Feynman-Bijl formula~\cite{Feynman1953, Feynman1954, Feynman1954, Girvin1986}.  In this literature, low-lying modes are described by the variational wavefunction in the first quantized form,
\begin{align}
    \psi_{{\bm{k}}}=\frac{1}{L^{d/2}}\uc_{{\bm{k}}} \phi_0=\frac{1}{L^{d/2}}\sum_{{\bm{x}}} e^{i{\bm{k}} \cdot {\bm{x}}} \phi_0,
\end{align}
 where $\phi_0$ is the exact ground state wavefunction. The difference between this original formulation and our construction is that our dispersing mode is written in second quantized form, where $\uc_{{\bm{k}}}=\frac{1}{L^{d/2}}\sum_{{\bm{x}}} e^{i{\bm{k}} \cdot {\bm{x}}} \uc_{{\bm{x}}}$. In addition, we chose to describe excitations over a specific ground state of fixed charge $\Vert m \rAngle$ or a Krylov sector $\Vert \cK \rAngle$.  Carrying on with the Feynman-Bijl derivation, the variational estimate for the energy of density fluctuation excitations is given by 
\begin{align}
    \epsilon_{{\bm{k}}} = \frac{\langle \psi_{{\bm{k}}}|  H - E_0|\psi_{{\bm{k}}}\rangle }{\langle \psi_{{\bm{k}}}|\psi_{{\bm{k}}}\rangle} = \frac{f({\bm{k}})}{s({\bm{k}})} ,
\end{align}
where $E_0$ is the exact ground state energy (which we set to zero). $f({\bm{k}})$ is called the oscillator strength, which can be evaluated as
\begin{align}
    f(\bm{k}) = \frac{1}{2L^d} \langle \phi_0| \left[\uc_{\bm{k}}^\dagger , [H,\uc^\vdagger_{\bm{k}}]\right]|\phi_0\rangle,  
\end{align}
and $s({\bm{k}})$ is the static structure factor:
\begin{align}
    s({\bm{k}}) = \langle \psi_{{\bm{k}}}|\psi_{{\bm{k}}}\rangle = \frac{1}{L^d} \langle \phi_0| \uc_{{\bm{k}}}^\dagger \uc^\vdagger_{{\bm{k}}}|\phi_0\rangle. 
\end{align}
In the context of superfluid Helium~\cite{Feynman1953, Feynman1954}, the oscillator strength $f(\bk) \sim k^2$ while the structural factor $s(\bk) \sim k$, giving rise to the linear dispersion of the density fluctuation modes $E_\bk \sim k$. 

For our problems of interest, the oscillator strength under the presence of $m$-th multipole conservation symmetry is given as $f({{\bm{k}}}) \sim k^{2(m+1)}$ for short-range interactions, and $f({{\bm{k}}}) \sim k^{2(m+\alpha)-d}$ for long-range interactions falling off as $1/{r^\alpha}$ when $ \frac{d}{2} + 1 > \alpha > \frac{d}{2}$. On the other hand, the static structural factor is generically constant, as elaborated in the next section, unless there is a constraint on the magnitude of local multipole density fluctuations.

\section{Orthonormal Basis States} \label{app:Ortho}

In this section, we examine the orthonormality of constructed density fluctuation modes. Specifically, we examine the orthonormality of the excitations in a Krylov sector, $\cK$,   
\begin{align}
    \Vert \cK_{{\bm{k}}} \rAngle \equiv \frac{1}{ \sqrt{ {\cal N}^{\cK}_{{\bm{k}}} }} \uc_{{\bm{k}}} \Vert \cK\rAngle.
\end{align}
As mentioned in the body of this paper, orthonormality plays an essential role in the construction of our variational modes in two respects. First, if orthogonality breaks down such that $\Vert \cK_{\bm{k}} \rAngle$ has significant overlap with the ground state, our variational modes may display a gapless dispersion even when the spectrum of our effective Hamiltonian, $\hat{ {\cal H} }_{\cal L}$, is gapped. Next, as was explained in the section on constrained dynamics, the dispersion of $\Vert \cK_{\bm{k}} \rAngle$ may depend on the normalization by its static structure factor, ${\cal N}^{\cK}_{{\bm{k}}} $

We begin by discussing a concrete example where a failure in orthonormality would result in incorrectly predicted relaxation times. Consider a charge conserving effective Hamiltonian that has a gapped spectrum. Due to the finite spectral gap, $\Delta E$, the relaxation should occur in $\cO(1)$ time. However, it is possible for the expected energy of our collective modes to still yield a gapless, quadratic dispersion: $\lAngle m_{{\bm{k}}} \Vert \hat{ {\cal H} }_{\cal L} \Vert m_{{\bm{k}}} \rAngle \underset{k \rightarrow 0}{\sim} k^2$, which would predict diffusive transport. 

This situation can arise if $\Vert m_{{\bm{k}}} \rAngle$ is formed from the superposition of the ground state and a small portion of a gapped excitation, $\Vert e_1 \rAngle$. For example, we can consider the following imagined decomposition at small k:
\begin{align}\label{eq:orthogonal_fail}
    \Vert m_{{\bm{k}}} \rAngle & \underset{k \rightarrow 0}= \sqrt{1-|c{\bm{k}}|^2}\Vert m\rAngle + c|{\bm{k}}|\Vert e_1\rAngle\\
    \lAngle m_{{\bm{k}}}\Vert m_{{\bm{k}}'} \rAngle
    &\underset{k \rightarrow 0}{\approx} 1 - \frac{|c|^2}{2}|{\bm{k}} -  {\bm{k}}'|^2\\
    \lAngle m_{{\bm{k}}} \Vert \hat{ {\cal H} }_{\cal L} \Vert m_{{\bm{k}}} \rAngle & \underset{k \rightarrow 0}{\propto} k^2 \lAngle e_1 \Vert \hat{ {\cal H} }_{\cal L}\Vert e_1\rAngle \propto k^2 \Delta E.
\end{align}
Where c is some $\cO(1)$ constant, and $\Delta E$ is the energy gap associated with $\Vert e_1\rAngle$.

However, overlap with the ground state is not the only way for orthogonality to fail. When considering individual Krylov sectors, if $\mathcal{K}$ is translation invariant and its dimension is at least extensive in system size, momentum is well-defined, and orthogonality of variational states $\Vert \cK_{\bm{k}} \rAngle$ follows directly. However, in general, a Krylov subspace, ${\cal K}$, may not be translation symmetric, i.e., $T_{{\bm{r}}}^\vdagger \cK T_{{\bm{r}}}^\dagger \neq \cK$ because multipole conservation and translation symmetries do not commute. In this case, our variational mode $\Vert \cK_{\bm{k}} \rAngle$ will not be a momentum eigenstate, and orthogonality does not follow. In order to circumvent this issue, we consider the \emph{symmetrized} Krylov subspace, $\cK^\textrm{s}$, as the following:
\begin{align}
    \cK^\textrm{s} \equiv \bigoplus_{{\bm{r}}} T_{{\bm{r}}}^\vdagger \cK T_{{\bm{r}}}^\dagger.
\end{align}
We can thus define a new momentum eigenmode $\Vert \cK^\textrm{s}_{{\bm{k}}} \rAngle=\sum_{\bm{r}} T_{\bm{r}}\Vert \cK_{\bm{k}}\rAngle/L^{d/2}$. The translation invariance of the Lindbladian ensures that the modes $\Vert \cK_{\bm{k}}\rAngle$ have the same energy expectation value as that of the symmetrized space:
\begin{align}
    \lAngle \cK^s_{\bm{k}} \Vert \lind \Vert \cK^s_{\bm{k}} \rAngle=\sum_{\bm{r}} \frac{\lAngle \cK_{\bm{k}} \Vert T_{\bm{r}}^\dagger \hat{ {\cal H} }_{\cal L} T_{\bm{r}}\Vert \cK_{\bm{k}}\rAngle}{L^d}=\lAngle \cK_{\bm{k}} \Vert \hat{ {\cal H} }_{\cal L} \Vert \cK_{\bm{k}}\rAngle,
\end{align}
where we use the fact that Krylov sectors are preserved under the action of $\lind$, but not translation, so that $\lAngle \cK_{\bm{k}} \Vert T_{\bm{r}}^\dagger \hat{ {\cal H} }_{\cal L} T_{\bm{r}'}\Vert \cK_{\bm{k}}\rAngle \sim \delta_{\bm{r},\bm{r}'}$. As such, we must now interpret $\bm{k}$ in $\Vert \cK_{\bm{k}}\rAngle$ as a label for the eigenstate that is distinct from the momentum. However, since $\Vert \cK_{\bm{k}}\rAngle$ shares the same spectral properties as $\Vert \cK^\textrm{s}_{{\bm{k}}} \rAngle$, we may exploit the translation invariance of $\Vert \cK^\textrm{s}_{{\bm{k}}} \rAngle$ to derive the compact form  of Eq.(\textcolor{red}{4}) from the main text. Still, a similar expression should exist for $\Vert \cK_{{\bm{k}}} \rAngle$, with $\boldsymbol{k}$ effectively entering as a mere integration variable.

Finally, the dispersion of our variational modes depends on a normalizing structure factor. In general, this may be difficult to calculate explicitly; however, in a charge conserving system, it can be directly calculated as
\begin{align}\label{eq:charge_structure}
{\cal N}_{{\bm{k}}} &= \sum_{\bm{x},\bm{x}'} \frac{e^{i \bm{k}\cdot (\bm{x}' - \bm{x}) }}{L^d} \lAngle m \Vert \uc_{\bm{x}} \uc_{\bm{x}'} \Vert m \rAngle \nonumber\\
&=\sum_{\bm{x} =\bm{x}'} \frac{\lAngle m \Vert \uc_{\bm{x}}^2 \Vert m \rAngle}{L^d} + \sum_{\bm{x} \neq \bm{x}'} \frac{e^{i \bm{k}\cdot (\bm{x}' - \bm{x}) }}{L^d} \lAngle m \Vert \uc_{\bm{x}} \uc_{\bm{x}'} \Vert m\rAngle \nonumber \\
&=\lAngle m \Vert \uc_{\bm{x}_0}^2 \Vert m\rAngle - \lAngle m \Vert \uc_{\bm{x}_0} \uc_{\bm{x}_0 + \bm{a}} \Vert m \rAngle,
\end{align}
where $\bm{x}_0$ and $\bm{a} \neq 0$ are arbitrary vectors. Here, we employ the fact that $\Vert m \rAngle$ is the projection onto the sector of total charge $m$, and $\Vert m \rAngle$ has no notion of distance. More precisely, it is invariant under the permutation of local sites. Accordingly, correlations between charges at different sites are the same for any two sites that are distinct. 
Thus, for a charge conserving system, the static structure factor, as given above, is a constant, independent of $\bm{k}$.  Alternatively, if $\uc_{\bm{x}}$ can be written as a $m^{\mathrm{th}}$-order derivative, $\uc_{\bm{x}}=\p_{\bm{x}}^m \hat{e}_{\bm{m},\bm{x}}$, the structure factor becomes
\begin{align}
{\cal N}_{{\bm{k}}} &= \sum_{\bm{x},\bm{x}'} \frac{e^{i \bm{k}\cdot (\bm{x}' - \bm{x}) }}{L^d} \lAngle m \Vert \p_{\bm{x}}^m \hat{e}_{\bm{m},\bm{x}} \p_{\bm{x}'}^m \hat{e}_{\bm{m},\bm{x}'} \Vert m \rAngle \nonumber \\
& \propto k^{2p} \sum_{\bm{x},\bm{x}'} \frac{e^{i \bm{k}\cdot (\bm{x}' - \bm{x}) }}{L^d} \lAngle m \Vert  \hat{e}_{\bm{m},\bm{x}}  \hat{e}_{\bm{m},\bm{x}'} \Vert m \rAngle\nonumber\\
&=k^{2p}\lAngle m \Vert  \hat{e}_{\bm{m},\bm{k}}  \hat{e}_{\bm{m},-\bm{k}} \Vert m \rAngle.
\end{align}
Thus, if a system has bounded fluctuations of $p^{\mathrm{th}}$ order moments, as described by variables, $\hat{e}_{\bm{m},\bm{k}}$, the structure factor will scale as ${\cal N}_{{\bm{k}}} \sim k^{2p}$.

\section{Long-Range Interactions} \label{app:H}

In this section, we discuss the derivation of the low-energy dispersion of our Lindbladian induced effective Hamiltonians, $\hat{ {\cal H} }_{\cal L}$, with long-range interactions that respect charge, dipole, or higher moment symmetries. 

\subsection{Charge Conservation}

Recalling the expression in the main text in Eq.(\textcolor{red}{10}), the expected variational energy with charge conservation is
\begin{align} \label{eq:charge_conserv_LRI}
 \lAngle m_{{\bm{k}}} \Vert \hat{ {\cal H} }_{\cal L} \Vert m_{{\bm{k}}} \rAngle & \propto \sum_{{\bm{x}}, {\bm{x}}'} \frac{(1 - \cos{{\bm{k}} \cdot({\bm{x}}'-{\bm{x}})})}{|{\bm{x}}-{\bm{x}}'|^{2\alpha} } \nonumber\\ 
  & \,\,\,\,\,\,  \times \lAngle m_{{\bm{k}}} \Vert \left[\tilde{{\cal O}}_{{\bm{x}}, {\bm{x}}'},  \uc_{{\bm{x}}} \right]^\dagger \left[\tilde{{\cal O}}_{{\bm{x}}, {\bm{x}}'},  \uc_{{\bm{x}}} \right]\Vert m_{{\bm{k}}} \rAngle  \nonumber\\
  & \quad \propto \int d^d {\bm{r}} \frac{(1 - \cos{{\bm{k}} \cdot{\bm{r}}})}{|{\bm{r}}|^{2\alpha} } \nonumber \\
  & \propto \int d\Omega_{d-2} \int_{1}^\infty d r  \int_{1}^{-1}d u \frac{(1 - \cos{(uk r )})}{|r|^{2\alpha-d+1} }\nonumber \\
& \propto \int_{1}^\infty  \frac{d r}{|r|^{2\alpha-d + 1}} \left( 2 - \frac{\sin{k r} }{k r} \right) \nonumber \\
   &=\frac{ {}_1 F_2 \left( \frac{d}{2} - \alpha ; \frac{3}{2},  \frac{d}{2} + 1-\alpha; -\frac{k^2}{4} \right) - 2}{2\alpha - d}  \nonumber\\   
   & + \Gamma(-1-2\alpha+d)\cos{\left(\pi \alpha - \frac{d\pi}{2}\right)} |k|^{2\alpha - d}\nonumber \\
   & \underset{k \rightarrow 0}{\propto} \left(C_1(\alpha) |k|^{2\alpha-d} + C_2(\alpha)k^2 \right),
\end{align}
where we have performed the spatial integral for $d \geq 3$, making use of the form of the angular integral. However, the asymptotic scaling form in the last line of Eq.~\eqref{eq:charge_conserv_LRI} also holds for $d=1,2$. 
Above, we used the substitution $u=\cos{\theta}$ with $\theta$ the angle between ${\bm{k}}$ and ${\bm{r}}$, ${}_1 F_2(a;b_1,b_2;z)$ is a hypergeometric function, and $C_1(\alpha)$ and $C_2(\alpha)$ are $\cO(1)$ coefficients obtained from this function that are smooth except at $\alpha=d/2$ and $1+d/2$. At $\alpha\leq d/2$, the spatial integral above exhibits IR divergences scaling with $\log L$ at $\alpha = d/2$ and with $L^{(d-2\alpha)}$ at $\alpha < d/2$, where $L$ is the linear system size.
Such a divergence of the single mode dispersion would lead to ultra-fast relaxation in the thermodynamic limit. However, for physical systems, we should renormalize the resulting dispersion to be bounded. In order to do so, we have to rescale the interaction strength with the diverging expression, which we label by $C_0(\alpha \leq \frac{d}{2},L)  = \int_{|\boldsymbol{r}|<L} d^d {\bm{r}} \frac{(1 - \cos{{\bm{k}} \cdot{\bm{r}}})}{|{\bm{r}}|^{2\alpha} }$. This rescaling leads to a finite energy gap at low $k$, and thus, relaxation within an $\mathcal{O}(1)$ time when $\alpha < d/2$.

Additionally, in this derivation, we assumed that the expectation $\lAngle m_{{\bm{k}}} \Vert \big[\tilde{{\cal O}}_{{\bm{x}}, {\bm{x}}'},  \uc_{{\bm{x}}} \big]^\dagger \big[\tilde{{\cal O}}_{{\bm{x}}, {\bm{x}}'},  \uc_{{\bm{x}}} \big]\Vert m_{{\bm{k}}} \rAngle$ did not depend on the distance, ${\bm{x}} - {\bm{x}}'$. This is true due to the same special property of the position-space representation of the $\Vert m\rAngle$ state that allowed us to simplify \eqnref{eq:charge_structure}:  
The correlation of two local operators acting on $\Vert m\rAngle$ only depends on whether the operators are at the same or distinct sites because the state $\Vert m \rAngle$ is invariant under the permutation of local sites. 
This is the case for  $\lAngle m_{{\bm{k}}} \Vert \big[\tilde{{\cal O}}_{{\bm{x}}, {\bm{x}}'},  \uc_{{\bm{x}}} \big]^\dagger \big[\tilde{{\cal O}}_{{\bm{x}}, {\bm{x}}'},  \uc_{{\bm{x}}} \big]\Vert m_{{\bm{k}}} \rAngle$, and the quantity is independent of the distance ${\bm{x}} - {\bm{x}}'$.

\subsection{Dipole Conservation}

Next, we turn to the case of dipole conservation.  Here, we look at generic dipole hoppings of the form 
\begin{align}
   h_{{\bm{x}},{\bm{x}}',\bm{n}}=\frac{\hat{S}^+_{{\bm{x}}}\hat{S}^-_{{\bm{x}}+\bm{n}}\hat{S}^-_{{\bm{x}}'}\hat{S}^+_{{\bm{x}}'+\bm{n}} + \textrm{h.c.} }{|{\bm{x}}-{\bm{x}}'|^{\alpha_0}|\bm{n}|^{\alpha_1}}.
\end{align}
Accordingly, our effective Hamiltonian will be of the form $\hat{ {\cal H} }_{\cal L}=\sum_{{\bm{x}},{\bm{x}}',\bm{n}} {\cal O}_{{\bm{x}}, {\bm{x}}'}^\vdagger {\cal O}_{{\bm{x}}, {\bm{x}}',\bm{n}}^\dagger$.  With this, we return to calculate the commutator from the main ext in Eq.(\textcolor{red}{8}).
 \begin{align}
    &[{\cal O}_{{\bm{x}}, {\bm{x}}',\bm{n}}, \uc_{{\bm{k}}}]=\sum_{{\bm{y}} \in \sS_{{\bm{x}}}} e^{i{\bm{k}} \cdot {\bm{x}}} [{\cal O}_{{\bm{x}}, {\bm{x}}',\bm{n}},  e^{i{\bm{k}}\cdot ({\bm{y}}-{\bm{x}})}  \uc_{{\bm{y}}} ] \nonumber \\ 
    &=e^{i{\bm{k}} \cdot {\bm{x}}}\frac{(1 - e^{ik\bm{n}})(1 - e^{i{\bm{k}} \cdot({\bm{x}}'-{\bm{x}})})}{|{\bm{x}}-{\bm{x}}'|^{\alpha_0}|\bm{n}|^{\alpha_1}} \left[\tilde{{\cal O}}_{{\bm{x}}, {\bm{x}}',\bm{n}}, \uc_{{\bm{x}}} \right],
 \end{align}
where $\tilde{{\cal O}}_{{\bm{x}}, {\bm{x}}',\bm{n}}={\cal O}_{{\bm{x}}, {\bm{x}}',\bm{n}}|{\bm{x}}- {\bm{x}}'|^{\alpha_0}|\bm{n}|^{\alpha_1}$ moves the power-law variation to the commutator prefactor. Note that this prefactor now carries the dependence on the displacement, ${\bm{r}}={\bm{x}} - {\bm{x}}'$, and dipole size, $\bm{n}$.  Repeating the same analysis as in the charge conserving case,
\begin{align}
     \lAngle \cK_{{\bm{k}}} \Vert \hat{ {\cal H} }_{\cal L} \Vert \cK_{{\bm{k}}} \rAngle 
 &\propto   \sum_{{\bm{r}}} \frac{(1 - \cos{{\bm{k}} \cdot({\bm{r}})})}{|{\bm{r}}|^{2\alpha_0} } \sum_{\bm{n}}\frac{(1-\cos{{\bm{k}} \cdot \bm{n}})}{ |\bm{n}|^{2\alpha_1}}\nonumber\\
 & \propto \begin{cases}
			 C_0(\alpha_0,L)\int_{{\bm{n}}} \frac{(1 - \cos{{\bm{k}} \cdot{\bm{n}}})}{|{\bm{n}}|^{2\alpha_1} } , & \alpha_0 \leq \frac{d}{2}\\
             \int_{{\bm{r}}} \frac{(1 - \cos{{\bm{k}} \cdot{\bm{r}}})}{|{\bm{r}}|^{2\alpha_0} } \int_{\bm{n}}\frac{(1-\cos{{\bm{k}} \cdot \bm{n}})}{ |\bm{n}|^{2\alpha_1}} & \alpha_0 > \frac{d}{2},
		 \end{cases}
\end{align}
where $C_0(\alpha,L)$ is a constant diverging with system size as discussed below \eqnref{eq:charge_conserv_LRI}. 
Now, we proceed with the above integration for each range of $\alpha_0$:
\begin{align}
&(\alpha_0 \leq d/2):  \nonumber \\
&\quad E_\bk \propto  C_0(\alpha_0)\left(C_1(\alpha_1) |k|^{2\alpha_1-d} + C_2(\alpha_1)k^2 \right) \underset{\alpha_1 \rightarrow \infty}{\propto} k^2\nonumber\\
&(\alpha_0 > d/2):  \nonumber \\
&\quad E_\bk \propto  \left(C_1(\alpha_0) |k|^{2\alpha_0-d} + C_2(\alpha_0)k^2 \right) \nonumber \\
& \qquad \qquad \times \left(C_1(\alpha_1) |k|^{2\alpha_1-d} + C_2(\alpha_1)k^2 \right), \nonumber \\
& \quad \,\, \underset{\alpha_1 \rightarrow \infty }{\propto} \left(C_1(\alpha_0) |k|^{2\alpha_0+2-d} + C_1(\alpha_0)k^4 \right),
\end{align}
where $C_1(\alpha_{0/1})$ and $C_2(\alpha_{0/1})$ are the same as before.

When $\alpha_1 \rightarrow \infty$, the effective Hamiltonian can be understood as describing long-ranged hopping of 2-local dipoles, $D_{\bm{x}} \equiv S^+_{{\bm{x}}}S^-_{{\bm{x}} + \bm{1}}$, where $\bm{1}$ is a unit vector. This case was discussed in the main text, with the $\alpha$-dependent phase diagram for transport behavior shown in Fig.\textcolor{red}{2} in the main text. On the other hand, if both $\alpha_{1/0} \leq \frac{d}{2}$, we have diverging integrals for both $\bm{r}$ and $\bm{n}$, which results in finite-time relaxation after renormalizing the divergence. Aside from this fast relaxation, we identify four distinct regimes (note that the physics is symmetric under the exchange of $\alpha_0 \leftrightarrow \alpha_1$):
\begin{enumerate}
\item $\alpha_{0},\alpha_1 > \frac{d}{2} + 1$:

This regime contains the limiting case $\alpha_{0/1} \rightarrow \infty$, corresponding to local dipoles with nearest neighbor hopings, and yields a dispersion $E_{{\bm{k}}} \sim k^4$.

\item $\alpha_{1} > \frac{d}{2} + 1$ and $\frac{d}{2} < \alpha_0 < \frac{d}{2} + 1$:

Local dipoles with long-range hoppings result in a dispersion $E_{{\bm{k}}} \sim k^{2(\alpha_0 + 1)-d}$

\item $\alpha_{1} > \frac{d}{2} + 1$ and $\alpha_{0} < \frac{d}{2}$:

Hoppings of local dipoles become so long-ranged that local charge transport arises from individual dipole creation/annihilation terms, equivalent to conventional charge conservation. This yields a dispersion $E_{{\bm{k}}} \sim k^{2}$.

\item $\frac{d}{2}<\alpha_{0},\alpha_1 <\frac{d}{2} + 1$,

Large dipoles with long-range hoppings yield a dispersion of $E_{{\bm{k}}} \sim k^{(2\alpha_0 -d)+(2\alpha_1 -d)}$. This gives rise to a dynamical exponent $z \in (0,4)$.
\end{enumerate}

\subsection{Multipole Conservation}

The extension of this to systems conserving $\{Q^{(0)},...,Q^{(m)}\}$ multipoles is straightforward. Interactions are composed of hoppings that scale as $\prod_{p=0}^{m}\frac{1}{|{\bm{r}}_p|^{\alpha_p}}$, where $r_p$ indicates the hopping distance between $(m-p)$-th moment charges. Therefore, $\alpha_0, \dots, \alpha_{m-2}, \alpha_{m-1}, \alpha_m$ control the locality of $m$-th moment hoppings ($(m+1)$-th moment lengths), ..., quadrupole hoppings (octopole lengths), dipole hoppings (quadrupole lengths), and charge hoppings (dipole lengths), respectively. From the above derivation, when all $\alpha_p > \frac{d}{2}$, our variational modes produce a dispersion of
\begin{align}\label{eq:longrange_general}
    \lAngle \cK_{{\bm{k}}} \Vert \hat{ {\cal H} }_{\cal L} \Vert \cK_{{\bm{k}}} \rAngle \underset{k \rightarrow 0}{\propto} \prod_{p=0}^{m}\left(C_1(\alpha_p) |k|^{2\alpha_p-d} + C_2(\alpha_p)k^2 \right).
\end{align}
Whenever $\alpha_p < \frac{d}{2}$, the $p$-th term in this product is replaced by  $C_0(\alpha_p,L)$, effectively acting as a constant upon renormalization of the divergence. As in the previous case, we enumerate four regimes:
\begin{enumerate}
\item $\alpha_0, \dots, \alpha_{m} >\frac{d}{2} + 1$ 

Local $m$-th moment charges with nearest neighbor hopings yield a dispersion of $E_{{\bm{k}}} \sim k^{2(m+1)}$

\item $\alpha_{1}, \dots, \alpha_{m} >\frac{d}{2} + 1$ and $\frac{d}{2} <\alpha_{0} < \frac{d}{2}+1$,

Local $m$-th moment charges with long-range hoppings yield a dispersion of $E_{{\bm{k}}} \sim k^{2(\alpha_0 + m)-d}$

\item $\alpha_{1}, \dots, \alpha_{m} > \frac{d}{2} + 1$ and $\alpha_{0} < \frac{d}{2}$,

Local $m$-th moment charges with extensive hoppings yield a dispersion of $E_{{\bm{k}}} \sim k^{2m}$

\item $\frac{d}{2} < \alpha_{0},\dots, \alpha_m < \frac{d}{2}+ 1$,

Extended $m$-th moment charges with long-range hoppings yield a dispersion $E_{{\bm{k}}} \sim k^{\sum_{p=0}^{m}(2\alpha_p -d)}$. Since $0 < 2 \alpha_i - d< 2$, the dynamical exponent $z = \sum_p (2\alpha_p - d) \in (0,2(m+1))$ covers the entire range between finite-time relaxation and conventional multipole subdiffusion.
\end{enumerate}

In the main text, we focus on cases 1-3, where transport can be accounted for by local excitations, however, our method accounts for dynamics for all ranges of different $\alpha_p$, where case 4 corresponds to a particular example.

\section{Numerical Details}
In the main text, we derived diffusive relaxation in dipole-conserving systems within Krylov sectors satisfying a charge area law (see Eq.(\textcolor{red}{12}) in the main text). In order to verify this prediction, we numerically evaluated the relaxation of classical systems exhibiting these constraints under discrete random time evolution. We emphasize that due to the universality of hydrodynamic transport, the same qualitative relaxation behavior is expected in thermalizing quantum many-body systems, see also Refs.~\cite{Iaconis19,morningstar2020kinetically,feldmeier2020anomalous,iaconis2021_multipole} for related approaches.

\begin{figure}[!t]
\centering
\includegraphics[width=\columnwidth]{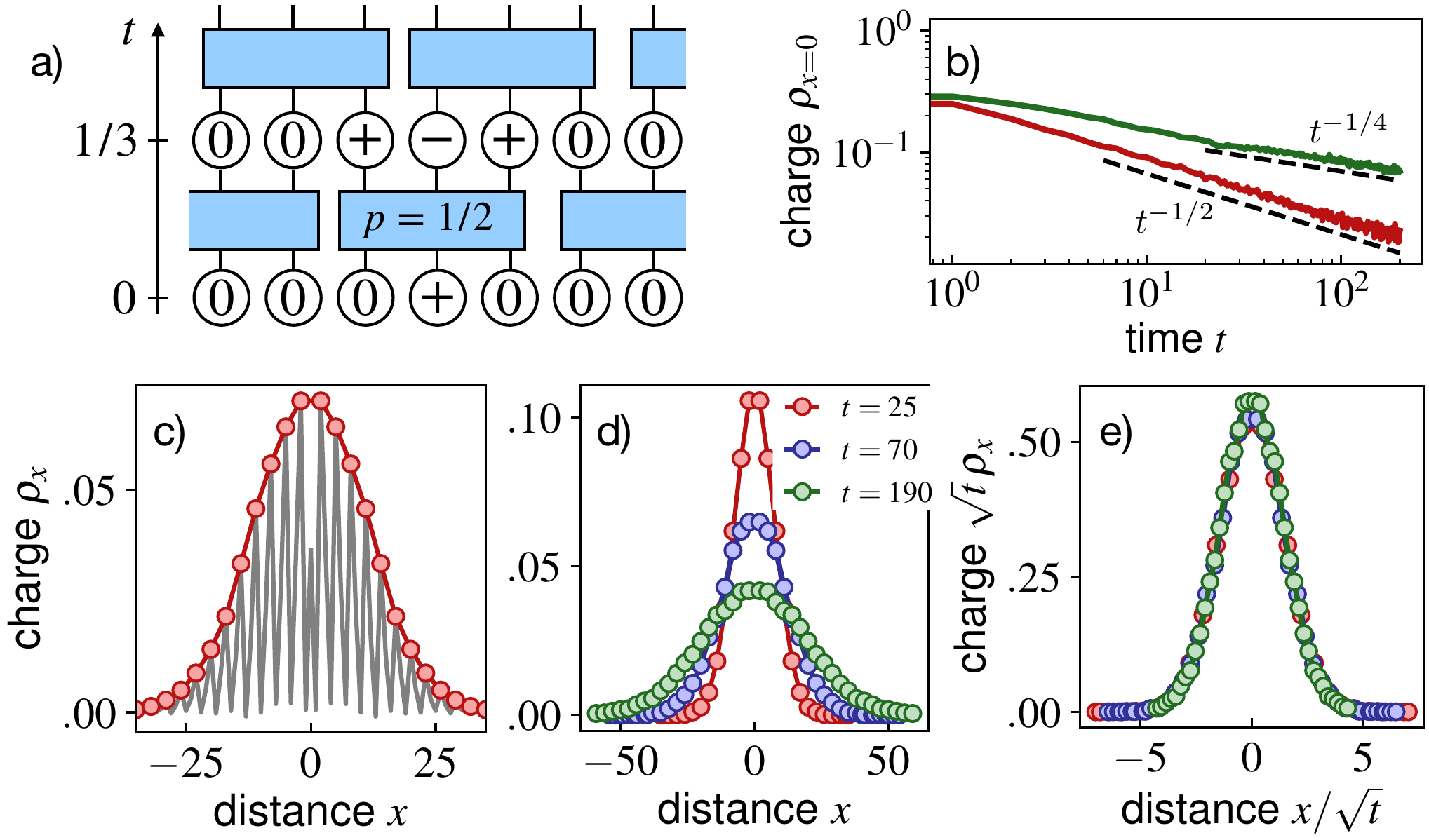}
\caption{\textbf{Numerical simulation of dipole-conserving dynamics.} \textbf{a)} We simulate the relaxation dynamics of a classical, discrete random time evolution, in which dipole-conserving updates of a given spatial range are performed randomly. \textbf{b)} For evolution with $3$-site updates, the charge excitation of the initial state shown in a) decays diffusively as $t^{-1/2}$ (red curve). In contrast, dynamics under $4$-site updates lead to subdiffusive decay $t^{-1/4}$ expected for generic systems (green curve). \textbf{c)} Profile $\uc(x,t)$ of the charge density at time $t=60$ of the evolution defined in a) with $3$-site updates. The red curve corresponds to an enveloping function. \textbf{d)} Enveloping functions of the charge density at different times. \textbf{e)} Diffusive scaling collapse of the enveloping functions shown in b). Numerical results were averaged over $2\times 10^5$ runs of the random time evolution in a chain of length $L=1000$.}
\label{fig:dipole}
\end{figure}

As the first example described in the main text, we time evolve the initial state $\ket{\psi_0}=\ket{...00+00...}$ of a $S=1$ spin chain with local Hilbert space $\uc_x \in \{0,\pm\}$. The time evolution is performed by applying local three-site updates that conserve both the charge $\sum_x \uc_x$ and dipole moment $\sum_x x\, \uc_x$ of the local three-site configuration. These updates are arranged in a brickwall pattern as depicted in \figref{fig:dipole}{a}. Furthermore, the updates are random, i.e. the updated charge configuration on the three sites is chosen randomly from all configurations within the same three-site charge and dipole sector.
In \figref{fig:dipole}{b}, we show the time evolution of $\mathbb{E}\left<\uc_{x=L/2}(t)\right>\sim t^{-1/2}$, confirming diffusive behavior. This is contrasted with the generic, subdiffusive decay $\sim t^{-1/4}$ seen when evolving the same initial state with similar local \textit{four-site} updates, for which the associated Krylov sector no longer follows a charge area law. In \figref{fig:dipole}{e}, we provide a scaling collapse of the full spatial profile of $\mathbb{E}\left<\uc_{x}(t)\right>$, again in agreement with diffusion. We note that the diffusive behavior in this Krylov sector has previously been reported in Ref.~\cite{skinner2022fracton}. Within the framework developed in our work, the emergence of diffusion in dipole conserving systems is explained as a consequence of the more general charge area-law constraint Eq.(\textcolor{red}{12}), from the main text.

To illustrate the generality of this result, in the main text we extended our analysis to systems beyond one spatial dimension. As a concrete example, we studied the dynamics of a two-dimensional dimer-vacancy model, subject to the hard-core constraint of maximally one dimer attached to each lattice site, see Fig.\textcolor{red}{3}a of the main text. Vacancies, i.e. sites without an attached dimer, carry a charge $\uc(\boldsymbol{x})=(-1)^{x_1+x_2}$. The constraint of either zero or one dimer on each bond of the lattice is equivalent to a finite electric field state space in a $U(1)$ link model formulation of this system~\cite{fradkin1990short,moessner_rvb}, ensuring area law charge fluctuations. We then explicitly demand the conservation of both the total charge and dipole moment under time evolution.

To verify the emergence of diffusive relaxation in this setup, we implemented a random, discrete classical time evolution similar to the previous example. Here, as depicted schematically in Fig.\textcolor{red}{3}a of the main text, we first apply plaquette updates which randomly update the state of elementary square plaquettes that contain either a horizontal or vertical pair of parallel dimers, or four charges. This process conserves the total charge and dipole moment. Then, we apply random dipole hoppings of neighboring $\pm$-charge pairs. Note that due to the sublattice structure of the local charge density $\uc_x$, conservation of the dipole moment implies that these charge pairs may only hop along displacement vectors $\boldsymbol{r}=(r_x,r_y)$ that satisfy $(r_x+r_y) \mod 2 = 0$. In our numerical implementation, we take into account $(|r_x|=2,r_y=0)$, $(r_x=0,|r_y|=2)$, and $(|r_x|=1,|r_y|=1)$.

\end{document}